\documentclass[11pt,twocolumn,twoside]{IEEEtran}
\usepackage{amsmath,amssymb}
\usepackage{tikz, pgfplots}
\graphicspath{ {./images/} }
\pgfplotsset{compat=1.5}
\usetikzlibrary{decorations.pathreplacing, decorations.pathmorphing, scopes,  shapes.misc, decorations.shapes, decorations.markings}
\usepgfplotslibrary{colormaps}
\usetikzlibrary{calc,external}
\tikzset{external/system call={lualatex \tikzexternalcheckshellescape -halt-on-error -interaction=batchmode -jobname "\image" "\texsource"}}
\colorlet{mycolor}{violet!40}

\newlength\figureheight
\newlength\figurewidth
\pgfplotsset{every linear axis/.append style={
    colorbar,
    colormap/jet,
    point meta min=0,
    xtick={0, 0.333, 0.666, 1, 1.333, 1.6666, 2},%
    xticklabels={0, $\frac{1}{3}$, $\frac23$,1, $1\frac13$, $1\frac23$, 2},
    enlargelimits=false}}
\pgfplotsset{/pgfplots/filter discard warning=false}
\pgfplotsset{every colorbar/.append style={scaled ticks=false, yticklabel style={/pgf/number format/.cd, fixed}}}
\def\norm#1{\|  #1 \|}

\def\abs#1{| #1 |}

\def\set#1{\{ #1 \}}
\def\ip#1#2{\langle #1 , #2 \rangle}

\def\supp{\mathop{\textstyle{\rm supp}}\nolimits}

\def\CHI{\hbox{\raise .5ex \hbox{$\chi$}}}

\def\det{\operatorname{det}}

\def\R{{\mathbb {R}}}
\def\C{{\mathbb {C}}}
\def\Z{{\mathbb {Z}}}
\def\N{{\mathbb {N}}}
\def\tfR{\R^2}
\def\choo#1#2{
\left(\!{\begin{smallmatrix}
 #1 \\ \! \! #2
\end{smallmatrix}}\!\right)
}

\newtheorem{theorem}{Theorem}[section]
\newtheorem{lemma}[theorem]{Lemma}

\newtheorem{definition}[theorem]{Definition}
\newtheorem{remark}[theorem]{Remark}
\newtheorem{example}[theorem]{Example}

\newsavebox{\activebox}

\title{Sampling and reconstruction of operators}

\author{G\"otz E. Pfander,~\IEEEmembership{Member,~IEEE,
}  and David Walnut
\thanks{Manuscript received XXX; revised YYY. This work was supported by the {\it Deutsche Forschungsgemeinschaft} (DFG) under grant
50292 DFG PF-4 Sampling Operators.}%
\thanks{G.E. Pfander is with Jacobs University, Germany and D. Walnut is with George Mason University.}}

\begin{document}
\maketitle

%
%

\begin{abstract}
We study the recovery of operators with bandlimited Kohn-Nirenberg symbol from the action of such operators on a weighted impulse train,
a procedure we refer to as operator sampling.  Kailath, and later Kozek and the authors have shown that operator sampling is possible
if the symbol of the operator is bandlimited to a set with area less than one.  In this paper we develop explicit reconstruction formulas
for operator sampling that generalize reconstruction formulas for bandlimited functions.  We give necessary and sufficient
conditions on the sampling rate that depend on size and geometry of the bandlimiting set.  Moreover, we show that under mild geometric
conditions, classes of operators  bandlimited to an unknown set of area less than one-half permit sampling and reconstruction.
A similar result considering unknown sets of area less than one  was independently  achieved  by Heckel and Boelcskei.

Operators with bandlimited symbols have been used to model doubly dispersive communication channels with slowly-time-varying impulse
response.  The results in this paper are rooted in work by Bello and Kailath in the 1960s.
\end{abstract}


\tableofcontents
%
%

\section{Introduction}

In this paper we develop a sampling theory and reconstruction formulas for operators bandlimited to domains of small area.
Analogously to the classical sampling theory of functions, the objective of operator sampling is to
fully characterize an object from at first sight insufficient information, specifically by observing an operator's action on a
single input, typically a discretely supported distribution, viz; a weighted delta train.
The theory developed herein applies to so-called bandlimited operators, defined as operators whose Kohn-Nirenberg symbol is bandlimited.
The symplectic Fourier transform of the Kohn-Nirenberg symbol of an operator is referred to as its spreading function, so that we
are considering operators whose spreading function is compactly supported or is contained in the fundamental domain of a lattice.
In engineering terms, the operators considered are characterized by limited time-frequency dispersion.

\subsection{Identification and sampling of operators}\label{sec:identificationandsampling}

The operator identification problem addresses the question whether an operator from a given class can be recovered from its action on a single probing
signal. That is, for a given class of operators $\mathcal H$, does there exist an input signal $g$ so that  $Hg$ determines $H$. Mathematically speaking,
we require that the map  $\Phi_g: \ H\mapsto Hg$ be injective on $\mathcal H$.  In order to be stable under noise introduced, for example, by physical
considerations or digital processing, it is reasonable to require in addition that the map $\Phi_g$ have a bounded inverse \cite{KP06}.

\begin{definition}
Let $\mathcal H$ be a collection of linear operators mapping a space of functions or distributions $X(\R)$ to a normed function space $Y(\R)$.
If for some $g\in X(\R)$,
$$\Phi_g: \mathcal H \longrightarrow Y(\R), \ H\mapsto Hg $$ is bounded above and below,
that is, if there are constants $0<A\le B<\infty$ such that
\begin{align}
 \label{eqn:identifyLINEAR}
 A \|H\|_{\mathcal{H}} \leq \|Hg\|_Y \leq B\,\|H\|_{\mathcal{H}}
\end{align}
for all $H\in\mathcal{H}$,
then we say that $\mathcal{H}$ is {\em identifiable with identifier $g\in X(\R)$}.
If $\mathcal H$ is not linear, then condition \eqref{eqn:identifyLINEAR} is replaced by
\begin{align}
 \label{eqn:identifyNONLINEAR}
 A \|H_1-H_2\|_{\mathcal{H}} \leq \|H_1g-H_2g\|_Y \leq B\,\|H_1-H_2\|_{\mathcal{H}}
\end{align}
for all $H_1,H_2\in\mathcal{H}$.
\end{definition}

We refer to operator identification as {\em operator sampling} when the identifier is a discretely supported distribution.


\begin{definition}\label{def:samplingset}
A strictly increasing sequence $\Lambda=\{\lambda_n\}_{n\in\Z}$ in $\R$ is a {\em set of sampling} for an operator class $\mathcal{H}$, if for some
never-vanishing sequence
$(d_n)_{n\in\Z}$, we have that $\sum_{n\in\Z} d_n \delta_{\lambda_n}$ identifies $\mathcal{H}$.  We define the {\em sampling rate} of $\Lambda$ by
$$D(\Lambda) = \lim_{r\to \infty} \frac{n^-(r)}{r}$$
where
$$n^-(r) = \inf_{x\in\R}\#\set{n\colon \lambda_n\in[x,x+r]}$$
assuming that the limit exists.
\end{definition}

\begin{remark}\label{rem:samplingset}
(1) $D(\Lambda)$ can be interpreted as the average number of  deltas
appearing in the identifier per unit time and corresponds to the {\em lower Beurling density} of $\Lambda$.

\noindent (2)
The assumption that the sequence $(d_n)$ never vanishes ensures that the sampling rate depends only on $\Lambda$.  In particular, we avoid
the situation in which for some set $\Lambda'\supseteq\Lambda$,
of higher density than $\Lambda$, $\sum_m d'_m\delta_{\lambda'_m} = \sum_n d_n\delta_{\lambda_n}$ where $d'_m=d_n$ whenever $\lambda'_m=\lambda_n$ and
$d'_m=0$ otherwise.
\end{remark}

In this paper we will consider mostly sampling sets that are periodic subsets of a fixed lattice on $\R$.

\begin{definition}\label{def:rationalsampling}
We say that an operator class $\mathcal{H}$ can be identified by {\em regular operator sampling} if there exists $T>0$, $L\in\N$, and a period-$L$
sequence $c=(c_n)$ such that $\sum_{n\in\Z} c_n\delta_{nT}$ identifies $\mathcal{H}$. \end{definition}

 In regular operator sampling, $D(\Lambda)=\norm{c}_0/(TL)$ where
$$\norm{c}_0 = \#\set{n\colon \mbox{$0\le n\le L{-}1$ and $c_n\ne 0$}}$$
is the support size of the vector $(c_0,\,\dots,\, c_{L-1})$.  In the remainder of this paper we will abuse notation
and not distinguish the vector $c\in\C^L$ from the doubly-infinite $L$-periodization $c=(c_n)$.

Our work addresses the identifiability of classes of operators characterized by their
Kohn-Nirenberg symbol being bandlimited to a set $S$ (the {\em spreading support}).

\subsection{Operator representations, bandlimited operators, and operator Paley-Wiener spaces}

Similarly to linear operators on finite dimensional space being represented by matrices,
the Schwartz kernel theorem implies that linear operators on any of the classical function
spaces on $\R$ can be represented by their kernel, that is, formally, we have
\begin{align}
  Hf(x)=\int \kappa_H(x,y) f(y)\, dy, \label{eqn:SchwartzKernel}
\end{align}
for a unique kernel $\kappa_H$.\footnote{In fact, with $\mathcal S (\R^d)$ denoting the space of Schwartz class functions and
$\mathcal S' (\R^d)$ its dual, we can associate to any linear and continuous operator mapping $\mathcal S (\R^d)$ to $\mathcal S' (\R^d)$
a kernel $\kappa\in \mathcal S' (\R^{2d})$ so that \eqref{eqn:SchwartzKernel} holds in a weak sense. Below, we shall consider
operators acting boundedly on the space of square integrable functions $L^2(\R)$ which fall in the framework outlined above.
We refer to \cite{Pfa10} for a more detailed functional analytic treatment of operator and function spaces involved.}

As operators are in 1-1 correspondence with their kernels, they can also be formally represented by their time-varying impulse response $h$,
their Kohn-Nirenberg symbol $\sigma$, or their spreading function $\eta$. In fact, formally,
\begin{align}
  Hf(x) &= \int h_H(x,t)\,f(x-t)\,dt \label{eqn:operatorrepresentations0} \\
  &= \iint \eta_H(t,\nu)\,e^{2\pi i\nu (x-t)}\,f(x-t)\,d\nu\,dt\label{eqn:operatorrepresentations1} \\
  &= \int \sigma_H(x,\xi)\, e^{2\pi i x \xi} \widehat f(\xi)\,d\xi,\label{eqn:operatorrepresentations2}
\end{align}
where
\begin{align}
h_H(x,t) &= \kappa_H(x,x-t) \nonumber  \\
    &= \int \sigma_H (x,\xi)\, e^{2\pi i \xi t}\, d\xi, \nonumber \\
    &= \int \eta_H(t,\nu)\, e^{2\pi i \nu (x-t)}\, d\nu \label{eqn:symbolrelations}
\end{align}
and the Fourier transform in \eqref{eqn:operatorrepresentations2} is normalized as $\mathcal F f (\xi)=\widehat f(\xi)=\int f(x)\, e^{-2\pi i x\xi}\,dx$.

Operator representations such as those given in \eqref{eqn:operatorrepresentations0}, \eqref{eqn:operatorrepresentations1},
\eqref{eqn:operatorrepresentations2} are considered in the theory of {\em pseudodifferential operators} where we write
$$
    \sigma(x,D)f(x)=\int \sigma(x,\xi)\, e^{2\pi i x \xi} \widehat f(\xi)\,d\xi.
$$
With the {\em symplectic Fourier transform} given by
$$
    \mathcal F_s F(t,\nu)= \iint F(x,\xi)\, e^{-2\pi i (x\nu -t\xi)}\,dx\,d\xi,
$$
\eqref{eqn:symbolrelations} implies $e^{-2\pi i t\nu }\,\eta_H (t,\nu)=\mathcal F_s \sigma_H (t,\nu)$.  We say that the operator
$H$ is {\em bandlimited} to the set $S\subseteq \R^2$ if $\supp \eta_H=\supp \mathcal F_s\sigma_H \subseteq S$.

Considering now spaces of such operators we arrive at the following definition.
\begin{definition}\label{def:operatorpaleywienerspaces}
 Given a set $S\subseteq\R^2$, define the {\em operator Paley-Wiener space} $OPW(S)$ by
\begin{align*}
OPW(S) &= \{H\in \mathcal L (L^2(\R), L^2(\R))\colon \\
       & \hskip.25in \supp \mathcal F_s \sigma_H= \supp\eta_H\subseteq S\}
\end{align*}
where $\mathcal L (L^2(\R), L^2(\R))$ denotes bounded operators on  $L^2(\R)$. The space of Hilbert-Schmidt operators  in $OPW(S)$ is
\begin{align*}
OPW^2 (S) &= OPW(S)\cap HS(L^2(\R)) \\
    &= \set{H\in \mathcal L (L^2(\R), L^2(\R))\colon \\
    & \hskip.25in \supp \mathcal F_s \sigma_H\subseteq S,\, \sigma_H\in L^2(\R^2)}.
\end{align*}
\end{definition}
The reconstruction formulas presented in this paper for $OPW^2 (S)$ hold formally for all of  $OPW(S)$.
Operator Paley-Wiener spaces defined by membership of the symbol in generic mixed $L^p$ spaces is considered in \cite{Pfa10};
see also Section~\ref{section:classicalsampling} below for some examples.

\subsection{Bandwidth of operators and analogies to classical sampling of functions}

The terminology {\em operator sampling} is intentionally suggestive of the classical theory of sampling of bandlimited functions,
and is justified for the following reasons.

\smallskip\noindent (1)  Classical sampling is in fact a special case of operator sampling in the following sense.  If for some fixed $\Omega>0$,
the operator class ${\cal H}$ consists of operators given by multiplication by functions $m\in L^2$ with $\supp\widehat{m}\subseteq[-\Omega/2,\Omega/2]$,
then choosing $0<T<1/\Omega$ and $g=\sum_n\delta_{nT}$, we have that for $H\in{\cal H}$ corresponding to multiplier $m$, $Hg=\sum_n m(nT)\delta_{nT}$
from which $m$ and hence $H$ can be recovered.  In this case, our general reconstruction formula \eqref{eqn:reconstructionformula} reproduces
the classical reconstruction formula.  For details, see Section~\ref{section:classicalsampling}.

Finally note that in this case, since $\norm{c}_0=L=1$, the (operator) sampling rate $\norm{c}_0/(TL)=1/T$ coincides with the sampling
rate in the classical sense.

\smallskip\noindent (2) In analogy with classical sampling, we can give a necessary condition on the (operator) sampling rate based on a
natural measure of the {\em bandwidth} of an operator modeling a time-varying communication channel.  T.~Kailath in \cite{Kai59}
defined the {\em bandwidth} of a channel with spreading function $\eta(t,\nu)$ by
$$\inf\set{B>0\colon \eta(t,\nu)=0,\, \forall t\in\R, \nu\notin(-B/2,B/2)}.$$
Taking into account possible gaps in the spreading support $S$, we can more precisely define the bandwidth by
\begin{equation}\label{eqn:bandwidth}
B(S) = \sup_{t\in\R}|\supp\eta(t,\nu)|  = \Big\|\int_\R \chi_{S}(\cdot,\nu)\,d\nu\Big\|_\infty
\end{equation}
where $\chi_S$ is the characteristic function of $S$.  This quantity can be interpreted as the maximum vertical extent of $S$.
The following theorem gives a necessary condition on a set
of sampling for the operator class with spreading support $S$.

\begin{theorem}\label{thm:necessarysamplingrate}
If $S$ is closed and $\Lambda$ is a set of sampling for $OPW^2(S)$ with $\inf \{ |\lambda - \mu|\,\colon \lambda,\mu\in \Lambda \}>0$,  then
$$D(\Lambda) \ge B(S).
$$
\end{theorem}

\smallskip\noindent (3) A sufficient condition on the (operator) sampling rate is more elusive and is tied to both
the area of the spreading support $S$ and its shape. However, if $|S|$ is small, then it suffices to observe  $Hg(t)$ on a
correspondingly small portion of the real line. For details, see Section~\ref{section:samplingrate}.

\smallskip\noindent (4) It should be noted that not only is classical sampling a special case of operator sampling, but also the
well-known result that time-invariant operators are characterized by their response to a delta centered at the origin.  Here $\Lambda=\{0\}$
and $S$ is a subset of the $t$-axis with $B(S)=0$.  For details, see Section~\ref{sec:classicalsampling2}

\subsection{Physical relevance of bandlimited operators}

In communications engineering, \eqref{eqn:operatorrepresentations0} and \eqref{eqn:operatorrepresentations1} are commonly used as models for linear (time-varying) communication channels.
The {\em time-varying impulse response} of the channel $h_H(x,t)$ is interpreted
as the response of the channel at time $x$ to a unit impulse at time $x-t$,
that is, originating $t$ time units earlier. Hence, if $h_H(x,t)\neq 0$ only for $0\leq t \leq T$, then $H$ is causal with maximum time-dispersion $T$.

If $h_H(x,t)=h_H(t)$ then the characteristics of the channel are time-invariant and in this case the
channel is a convolution operator.  As mentioned above, such channels are identifiable since $h_H(t)$ is the response of the channel
to the input signal $\delta_0(t)$, the unit-impulse at $t=0$.

A mobile communication channel has the property that $h_H(x,t)$ depends on $x$, but changes as a function of $x$ rather slowly,
since the change in the channel, for example, by movement of receiver, transmitter, or reflecting objects, is slow when compared
with the speed of light at which information travels. This slow variance can be expressed through a bandlimitation of $h_H(x,t)$
as a function of $x$, that is, as a support constraint on
the {\em spreading function}  of $H$, $\eta_H(t,\nu) = \int h_H(x,t)\,e^{-2\pi i\nu (x-t)}\,dx$, as a function
of $\nu$. We conclude that a  causal doubly dispersive communications channel with maximum time dispersion $T$, and $h_H(x,t)$
bandlimited in $x$ to $[-\Omega/2, \Omega/2]$ is represented by a spreading function supported on  the set $[0,T]{\times}[-\Omega/2, \Omega/2]$,
that is, by operators in $OPW^2([0,T]{\times} [-\Omega/2, \Omega/2])$ since $|\eta_H| =|\mathcal F_s \sigma_H|$.

To substantiate this bandlimitation on $\sigma_H(x,t)$ further, we denote translation by $t$ by $T_t: f(x)\mapsto f(x-t)$ and modulation by $\nu$ by
$M_\nu: f(x)\mapsto e^{2\pi i\nu x}\,f(x)$.
The latter is also referred to as frequency shift as $\widehat{M_\nu f}=T_\nu\, \widehat f$.
Then \eqref{eqn:operatorrepresentations1} becomes the operator-valued integral
\begin{align*}
H
&  =    \int\!\!\!\!\int \eta_H(t,\nu)\,T_t\,M_\nu\,d\nu\,dt \\
&  = \int_0^T\!\!\int_{-\Omega}^{\Omega} \eta_H(t,\nu)\,T_t\,M_\nu\,d\nu\,dt\,,
\end{align*}
that is, the spreading function is the coefficient vector of the time-frequency shifts that a communication channel carries out.
Hence, $OPW^2([0,T]{\times} [-\Omega/2, \Omega/2])$ has {\em maximum time-delay} $T$ and {\em maximum frequency spread} $\Omega$.

\subsection{Relation to other work}

In 1959, T. Kailath \cite{Kai59, Kai62, Kai63} asserted that for time-variant communication
channels to be identifiable it is necessary and sufficient that the maximum time-delay, $a$, and Doppler spread, $b$, satisfy
$ab\le 1$ and gave a convincing justification for his assertion on signal-theoretic grounds.
Kailath considers the response of the channel to a train of impulses separated by at least $a$ time units, so that in this sense
the channel is being ``sampled'' by a succession of evenly-spaced impulse responses.  The condition $ab\le 1$ allows for
the recovery of sufficiently many samples of $h_H(x,t)$ to determine it uniquely.  To prove necessity, Kailath assumes that the
channel can be identified by a probing signal, $g$, essentially both time- and band-limited.  If the response, $Hg$, is also so limited,
the number of degrees of freedom in $Hg$ can be estimated.  This number is then compared to the number of degrees of freedom in the
impulse response $h_H(x,t)$ under the same time and band-limiting assumption as on $g$ in each variable.  Comparing degrees of
freedom leads to the necessary inequality $ab\le 1$.

Kailath's assertion was given the precise mathematical framework described in Section~\ref{sec:identificationandsampling}
and proved in \cite{KP06}.

In 1969, P.~A.~Bello \cite{Bel69} argued that what is important
for channel identification is not the product $ab$ of the maximum time-delay and Doppler
shift of the channel but the area of the support of the spreading function.  It is notable that Kailath also asserted
something along these lines.
This means that a time-variant channel whose spreading function has essentially arbitrary support is identifiable
as long as the area of that support is smaller than one.
Using ideas from \cite{KP06}, Bello's result was confirmed in \cite{PW06b}.

Building on findings in \cite{KP06,PW06b,PW06}  a number of results have been established that are now part of the herein described
sampling theory for operators. For example, the results in  \cite{PW06b} were extended from the setting of Hilbert-Schmidt operators
to a much wider class of pseudodifferential operators in \cite{Pfa10}. In \cite{HP09}, the choice of non periodic (irregular/jitter)
sampling locations for operator sampling was discussed. Necessary and sufficient conditions for the identifiability of  bandlimited
Multiple Input Multiple Output (MIMO) channels were given in \cite{Pfa08}.

More recently, sampling results for stochastic operators, that is, for operators with stochastic spreading functions, have been obtained
\cite{OPZ11,PZ11a,PZ11b}.  Also, in applications, it is required to replace the identifier considered in this paper by finite time or
finite bandwidth, that is, smooth, signals.  Local recovery results in this setting, as well as a reconstruction formula that allows
for the application of coarse quantization methods prior to the approximate recovery of the operator are given in \cite{KP12}.
Focusing on a parametric setup, the identification of bandlimited operators was analyzed with respect to applicability in
super-resolution radar \cite{BGE11}.

In Section~\ref{sec:unknownsupport}, we address the problem of identifiability of operators with unknown bandlimitation.
Independently, Heckel and Boelcskei (\cite{BH11, BH13}) have obtained a result similar to Theorem~\ref{thm:unknownDomain}
characterizing identifiability of a certain
(nonlinear) class of operators with spreading support of area $\le 1/2$.  Theorem~\ref{thm:unknownDomain} gives a sufficient condition
for a more general class of operators, and Theorem~\ref{thm:characterizationunknownsupport} generalizes the result in \cite{BH11, BH13}.
In addition, Heckel and Boelcskei (\cite{BH11, BH13}) prove a remarkable result in which they prove
identifiability for unknown support sets of area less than one, rather than $\le 1/2$.

\section{Main results}

\subsection{Properties of Gabor system matrices}

The basic strategy for operator sampling described in this paper was laid out in \cite{PW06b}.  The idea is to translate
the reconstruction problem into an {\em a priori} under-determined linear system whose coefficients come from a finite
Gabor system, and then give conditions under which that system can be solved.
More specifically, given $H\in OPW^2(S)$, $T>0$, and $L\in\N$, let $g=\sum_n c_n\,\delta_{nT}$ for some period-$L$
sequence $c=(c_n)$.  Then from the response $Hg(x)$, we can derive the $L\times L^2$ linear system
\begin{equation}\label{eqn:basiclinearsystem}
{\bf Z}_{Hg}(t,\nu) = G(c)\,\boldsymbol\eta_H(t,\nu)
\end{equation}
where ${\bf Z}_{Hg}(t,\nu)$ is an $L$--vector computed directly from $Hg$, $\boldsymbol\eta_H(t,\nu)$ is an $L^2$--vector consisting
of shifts of a periodized version of the spreading function $\eta_H$ of $H$ (see Lemma~\ref{lem:matrixequationquasiperiodic}),
and $G(c)$ is an $L\times L^2$ Gabor system matrix defined as follows.

\begin{definition}\label{def:translationmodulation}
Given $L\in\N$, let $\omega=e^{2\pi i/L}$ and define the {\em translation operator $T$}
on $(x_0,\,\dots,\,x_{L-1})\in\C^L$ by
$$Tx=(x_{L-1},x_0,\,x_1,\,\ldots,x_{L-2}),$$
and the {\em modulation operator $M$} on $\C^L$ by
$$Mx=(\omega^0 x_0, \omega^1 x_1,\,\dots,\, \omega^{L-1} x_{L-1}).$$
Given a vector $c\in\C^L$ the {\em finite Gabor system with window $c$}
is the collection $\set{T^q M^p c}_{q,p=0}^{L-1}$.
Define the {\em full Gabor system matrix} $G(c)$
to be the $L\times L^2$ matrix
\begin{equation}\label{def:fullgabormatrix}
G(c) = \left[\,\, D_0\,W_L\,\,\vrule\,\, D_1\,W_L\,\, \vrule \,\,\cdots \,\, \vrule\,\, D_{L-1}\,W_L \,\,\right]
\end{equation}
where $D_k$ is the diagonal matrix with diagonal $T^kc = (c_{L-k},\,\dots,\,c_{L-1},\,c_0,\,\dots,\,c_{L-k-1})$,
and $W_L$ is the $L\times L$ Fourier matrix $W_L = (e^{2\pi inm/L})_{n,m=0}^{L-1}$.
\end{definition}

\begin{remark}\label{rem:translationmodulation}
(1) For $0\le q,\,p\le L-1$, the $(q+1)$st column of the submatrix $D_pW_L$ is the vector $M^pT^qc$ where the operators $M$ and $T$
are as in Definition~\ref{def:translationmodulation}.  This means that each column of the matrix $G(c)$
is a unimodular constant multiple of an element of the finite Gabor system with window $c$, namely
$\set{e^{-2\pi ipq/L}\,T^q M^pc}_{q,p=0}^{L-1}$.

\noindent (2)
Note that the finite Gabor system defined above consists of $L^2$ vectors in $\C^L$ which form an overcomplete
tight frame for $\C^L$ \cite{LPW05}.  For details on Gabor frames in finite dimensions, see \cite{LPW05,KPR08,FKL09}
and the overview article \cite{Pfa12}.
\end{remark}

The reconstruction formulas in this paper are based on explicitly and uniquely solving (\ref{eqn:basiclinearsystem}).
for this purpose we require conditions on $G(c)$ under which this is possible.

\begin{definition}\cite{DE03}
The {\em Spark} of an $M\times N$ matrix F is the size of the smallest linearly dependent
subset of columns, i.e.,
$$Spark(F) = \min\set{\norm{x}_0\colon Fx=0,\ \ x\ne 0}$$
If $Spark(F)=M+1$, then $F$ is said to have {\em full Spark}.
$Spark(F)=k$ implies that any collection of fewer than $k$ columns of $F$
is linearly independent.
\end{definition}

The existence of Gabor matrices with full Spark has been addressed in \cite{LPW05} and \cite{M13}.

\begin{theorem}\label{thm:lpw05}\cite{LPW05}
If $L$ is prime, then there exists a dense, open subset of $c\in\C^L$ such that every minor of $G(c)$ is nonzero.
In particular, for such $c$, $G(c)$ has full Spark.
\end{theorem}

Note that if $L$ is not prime then the result of this theorem does not hold.  That is, if $L$ is not prime, then for
any $c\in\C^L$ there is a minor of $G(c)$ that vanishes.  However, it has recently been shown by Malikiosis
that for any $L\in\N$, we can get the second half of the conclusion.

\begin{theorem}\label{thm:malikiosis}\cite{M13}
For every $L\in\N$ there exists a dense, open subset of $c\in\C^L$ such that $G(c)$ has full Spark.
\end{theorem}

This next result states that, again assuming that $L$ is prime, the Spark of the matrix $G(c)$
is related to the support size of the vector $c$.

\begin{theorem}\label{thm:smallsparksampta13}\cite{PW13}
If $L\in\N$ is prime, and $k\le L$, there exists $c\in\C^L$ with the property that $Spark(G(c))=k+1$,
and $\supp(c)\subseteq\set{0,\,1,\,\dots,\,k-1}$.  Moreover, the set of such $c$ forms an open, dense subset of $\C^k\times\{0\}$.
\end{theorem}

These theorems show that it is possible to choose a period-$L$ sequence $c$ such that the system (\ref{eqn:basiclinearsystem})
always has a solution as long as there are no more than $L$ non-vanishing unknowns on the right side.  In fact, if $L$ is prime, we can say a bit more,
namely that if there are no more than $k\le L$ non-vanishing unknowns on the right side, then we can guarantee solvability
with a window $c$ supported on no more than $k$ contiguous indices.

\subsection{Necessary and sufficient conditions for identifiability of $OPW^2(S)$}

In this section, we explore conditions under which the operator class $OPW^2(S)$ is identifiable.  We give necessary and sufficient
conditions on $S$ under which identification is possible with any identifier, then characterize when this is possible using regular operator
sampling.

In \cite{KP06,PW06b} (cf. \cite{PW06} and \cite{Pfa10}), the following result is given.
Here and in the following, $|S|$ denotes the Lebesgue measure of the set $S$.

\begin{theorem}\label{thm:OPWidentifiable}
$OPW^2(S)$ is identifiable by regular operator sampling if $S$ is compact and $\abs{S}<1$, and not
identifiable if $S$ is open and $\abs{S}>1$.
\end{theorem}

The following result guarantees the existence of a discretely supported identifier for support sets $S$ with $|S|\le 1$
that satisfy  certain periodization conditions.  The result characterizes operator Paley Wiener spaces  that can be identified by regular
operator sampling.

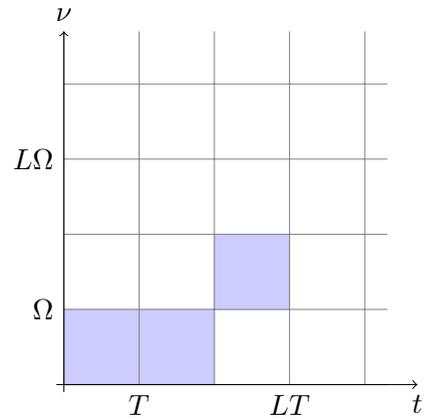
\begin{figure}
  \centering
\begin{tikzpicture}[scale=0.5]
\def\L{9}
\def\Lminusone{8}
\filldraw [fill=blue!20, draw=blue!20] (0,0) -- (0,2) -- (2,2) -- (2,0) -- (0,0);
\filldraw [fill=blue!20, draw=blue!20] (2,0) -- (2,2) -- (4,2) -- (4,0) -- (2,0);
\filldraw [fill=blue!20, draw=blue!20] (4,2) -- (4,4) -- (6,4) -- (6,2) -- (4,2);
%
%

\draw[style=help lines, step={(2,2)}] (0,0) grid (\L-0.4, \L+0.4);
\draw[->] (-0.2,0) -- (\L+0.4,0) node[below] {$t$};
\draw[->] (0,-0.2) -- (0,\L+0.4) node[above] {$\nu$};
\node[below] at (2,0) {$T$};
\node[left] at (0,2) {$\Omega$};
\node[below] at (6,0) {$LT$};
\node[left] at (0,6) {$L\Omega$};
\end{tikzpicture}
  \caption{The space $OPW^2(S)$ is identifiable for $S$ (in blue) with area 1 as it clearly satisfies
   \eqref{eqn:characterization1} and \eqref{eqn:characterization2}. $S$ has a $(T,3)$-rectification and $B(S)=\Omega$.
    Such sets were considered in \cite{KP06,PW06b,PW06}.
    Recovery of operators in $OPW^2(S)$ is possible using the reconstruction formula (\ref{eqn:reconstructionformulaperiodic}).}
  \label{fig:rectificationB}
\end{figure}

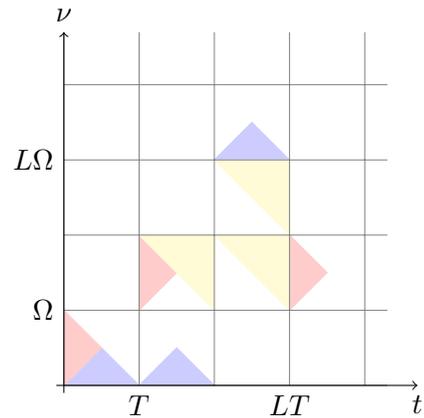
\begin{figure}
  \centering
\begin{tikzpicture}[scale=0.5]
\def\L{9}
\def\Lminusone{8}
\filldraw [fill=blue!20, draw=blue!20] (0,0) -- (1,1) -- (2,0) -- (0,0);
\filldraw [fill=blue!20, draw=blue!20] (2,0) -- (3,1) -- (4,0) -- (2,0);
\filldraw [fill=blue!20, draw=blue!20] (4,6) -- (5,7) -- (6,6) -- (4,6);

\filldraw [fill=red!20, draw=red!20] (0,0) -- (1,1) -- (0,2) -- (0,0);
\filldraw [fill=red!20, draw=red!20] (2,2) -- (3,3) -- (2,4) -- (2,2);
\filldraw [fill=red!20, draw=red!20] (6,2) -- (7,3) -- (6,4) -- (6,2);

\filldraw [fill=yellow!20, draw=yellow!20] (6,4) -- (4,6) -- (6,6) -- (6,4);
\filldraw [fill=yellow!20, draw=yellow!20] (4,2) -- (2,4) -- (4,4) -- (4,2);
\filldraw [fill=yellow!20, draw=yellow!20] (6,2) -- (4,4) -- (6,4) -- (6,2);

\draw[style=help lines, step={(2,2)}] (0,0) grid (\L-0.4, \L+0.4);
\draw[->] (-0.2,0) -- (\L+0.4,0) node[below] {$t$};
\draw[->] (0,-0.2) -- (0,\L+0.4) node[above] {$\nu$};
\node[below] at (2,0) {$T$};
\node[left] at (0,2) {$\Omega$};
\node[below] at (6,0) {$LT$};
\node[left] at (0,6) {$L\Omega$};
\end{tikzpicture}
  \caption{The union of the colored sets, $S$, satisfies  \eqref{eqn:characterization1} and \eqref{eqn:characterization2}.  Hence, $OPW^2(S)$  is identifiable by a weighted delta train with period-$3$ weighting sequence even though a $(T,3)$-rectification is not possible (note that $7>3$
  boxes are active). Recovering $\eta$ from $Hg$ using \eqref{eqn:basiclinearsystem} directly requires solving three systems of linear equations, one to recover $\eta$ on the yellow support set, one to recover $\eta$ on the red support set, and one to recover $\eta$ on the blue support set.
  $H\in OPW^2(S)$ can be reconstructed using formula \eqref{eqn:reconstructionformulaperiodic2}.
  Note also that $B(S)=2\Omega$ and that the sampling rate is $1/T=3\Omega>2\Omega$.  }
  \label{fig:rectification}
\end{figure}

%
%
%
\begin{theorem}\label{thm:characterization}
 Let $g=\sum_{n\in\Z} c_n \delta_{nT}$ with $c\in\C^L$ chosen so that $G(c)$ has full Spark.
 For $S\subseteq \R^2$  the following are equivalent.
\begin{enumerate}
\item[{\rm (i)}]
The map $\Phi_g: OPW^2(S) \rightarrow L^2(\mathbb{R}), \ H \mapsto Hg$ is injective.
 \item[{\rm (ii)}]
 The function $g$ identifies $OPW^2(S)$.
 \item [{\rm (iii)}]
 $S$ is a subset of a fundamental domain of the lattice $LT\Z \times (1/T) \Z$, that is,
 \begin{equation}\label{eqn:characterization1}
 \sum_{k,\ell} \chi_{S+(kLT,\ell/T)}\leq 1\quad a.e.
 \end{equation}
 and $S$ periodized by the lattice $T\Z \times 1/(TL) \Z$ is at most an $L$-cover, that is
 \begin{equation}\label{eqn:characterization2}
 \sum_{k,\ell} \chi_{S+(kT,\ell/(TL))}\leq L\quad a.e.
 \end{equation}
\end{enumerate}
\end{theorem}

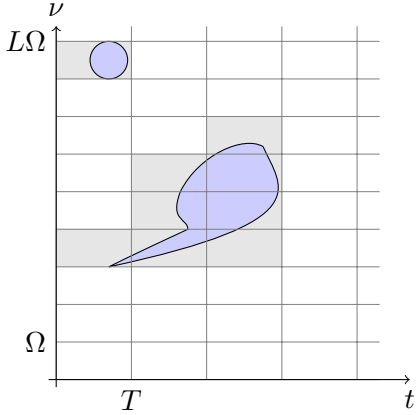
\begin{figure}[htbp]
  \centering
\begin{tikzpicture}[scale=0.5]
\def\L{9}
\def\Lminusone{8}
\begin{scope}[draw=gray!20, fill=gray!20]      
\filldraw (0,3) -- ++(0,1) -- ++(2,0) -- ++(0,2) -- ++(2,0) --  ++(0,1) -- ++(2,0) -- ++(0,-4)    ;
\filldraw (0,8) -- ++(0,1) -- ++(2,0) -- ++(0,-1)    ;
\end{scope}

\filldraw[fill=blue!20, draw=black]
                (1.4,3) .. controls(2, 3.3)  .. (3.5,4.0)
						.. controls (3.5,4.3) and (3.0,4.2) .. (3.3,5)
						.. controls (3.8,6) and (5, 6.5) .. (5.5,6.2)
						.. controls (6, 5.1) and (7, 4.2) .. (1.4,3)
						(1.4, 8.5) circle (.5);
\draw[style=help lines, step={(2,1)}] (0,0) grid (\L-0.4, \L+0.4);
\draw[->] (-0.2,0) -- (\L+0.4,0) node[below] {$t$};
\draw[->] (0,-0.2) -- (0,\L+0.4) node[above] {$\nu$};
\node[below] at (2,0) {$T$};
\node[left] at (0,1) {$\Omega$};
\node[left] at (0,9) {$L\Omega$};
\end{tikzpicture}
  \caption{The set $S$ in blue, its rectification in gray.  We have $L=9$ and $T\Omega=1/9$. }
  \label{fig:rectificationA}
\end{figure}

See Figures \ref{fig:rectificationB}-\ref{fig:rectificationA} for an illustration of spreading supports sets
$S$ that lead to identifiable operator Paley Wiener spaces.

\begin{remark}\label{rem:characterization}
(1) It is clear that if $S$ is bounded, then (\ref{eqn:characterization1}) is satisfied as soon as $S$ is contained in a
rectangle of width $TL$ and height $1/T$.

\noindent (2) Note that (\ref{eqn:characterization2}) implies that $|S|\le 1$, and that if $|S|=1$,
the cover must be an exact $L$-cover, that is,
$$ \sum_{k,\ell} \chi_{S+(kT,\ell/(TL))}= L\quad a.e.$$

\noindent (3) As discussed in detail in Remark~\ref{rem:rectification} below, for any compact set $S$ with $|S|< 1$, there exists $T$,
$L$ so that (\ref{eqn:characterization1}) and   (\ref{eqn:characterization2}) hold.

\noindent (4)
Note that (\ref{eqn:characterization1}) and   (\ref{eqn:characterization2}) are satisfied for some unbounded sets with area
less than or equal to one, for example,
$$S= \Big( \bigcup_{n=0}^{\infty} [n+1-2^{-n},n+1- 2^{-(n+1)}]\Big)\times [-\tfrac 1 2,\tfrac 1 2]$$
permits the choice of $T=1$ and $L=1$.

\noindent (5)  On the other hand, it is not hard to construct an unbounded set $S$ of arbitrarily small measure so that for all $T$ and $L$,
(\ref{eqn:characterization1}) fails.  Indeed, let $\{q_n\}_{n\in \mathbb N}$ be an enumeration of the countable
set of rational numbers $\mathbb Q$. For $\epsilon,\delta >0$ set
$$S_{\epsilon,\delta}= \Big( [-\epsilon,\epsilon]\cup \bigcup_{n=0}^{\infty} 2^{-n} [-\epsilon,\epsilon] + q_n\Big)\times [-\delta,\delta].$$
We have $|S_{\epsilon,\delta}|<8\epsilon\delta$ since we are taking the union of sets that are not disjoint, in fact,
every set in the union contains countably many sets in the union.  In order to show that  there exist no $T>0$ and $L\in\N$
such that (\ref{eqn:characterization1}) holds, observe first that clearly, $LT>\epsilon$, and there
exists $n_0\in\N$ so that $|q_{n_0}-LT|<\epsilon$. But then $S_{\epsilon,\delta}-LT$ intersects with $S_{\epsilon,\delta}$
on a set of positive measure since
$$\big|[-\epsilon,\epsilon]\cap 2^{-n} [-\epsilon,\epsilon] + q_{n_0}-LT\big|>0.$$

\noindent (6) If $Spark(G(c))=K\le L$, then $OPW^2(S)$ is identifiable if the upper bound $L$ in (\ref{eqn:characterization2})
is replaced by $K-1$.

\noindent (7) The conditions (\ref{eqn:characterization1}) and (\ref{eqn:characterization2})
are related to the {\em rectification} of the region $S$, that is, its efficient covering by small rectangles.
(See Fig~\ref{fig:rectificationA}).
\end{remark}

\begin{definition}\label{def:rectification}
Let $S\subseteq\tfR$, $|S|\le 1$, $T>0$, and $L\in\N$ be given.  We say that $S$ admits a {\em $(T,L)$-rectification} if
\begin{enumerate}
\item[{\rm (a)}] $S$ is contained in a fundamental domain of the lattice $(TL)\Z\times(1/T)\Z$, and
\item[{\rm (b)}] the set
\begin{equation}\label{eqn:periodizationofS}
S^\circ=\bigcup_{(k,\ell)\in\Z^2} S+(kTL,\ell/T)
\end{equation}
meets at most $L$ rectangles of the form $R_{q,m}=[0,T]\times[0,1/TL]+(qT,m/TL)$, $0\le q, m < L$.
The {\em active boxes} in the rectification are indexed by
$$\Gamma = \set{(q,m), 0\le q, m<L \colon  R_{q,m}\cap S^\circ \ne\emptyset}.$$
\end{enumerate}
\end{definition}

It is clear that \eqref{eqn:characterization1} and \eqref{eqn:characterization2} are
satisfied if $S$ admits a $(T,L)$-rectification, but Figure~\ref{fig:rectification} illustrates
that the converse is not true.  However, \eqref{eqn:characterization1} and \eqref{eqn:characterization2}
allow for the linear system \eqref{eqn:basiclinearsystem} to change depending on the point $(t,\nu)$.
In fact, such an observation further characterizes regions $S$ such that $OPW^2(S)$ can be identified by regular operator
sampling.

\begin{lemma}\label{lem:rectification3}
Suppose that for some $T>0$ and $L\in\N$, $S$ satisfies (\ref{eqn:characterization1}).  Then
$$ \sum_{k,\ell} \chi_{S+(kT,\ell/(TL))}\leq K\le L\quad a.e.$$
if and only if there exists a partition $\set{A_j}_{j=1}^N$ of the set $[0,T]\times[0,1/(LT)]$ with the property that
for each $j$, at most $K$ of the sets $A_j+(kT,\ell/(LT))$, $0\le k,\ell<L$ meet $S^\circ$.
Moreover, $S$ can be partitioned as
\begin{equation}\label{eqn:disjointunion}
S = \bigcup_{j=1}^N S_j
\end{equation}
where
$$S_j = S\cap\bigg[\bigcup_{k,\ell\in\Z} A_j+(kT,\ell/(LT))\bigg]$$
and where each $S_j$ admits a $(T,L)$-rectification with $\abs{\Gamma}\le K$.
\end{lemma}

\begin{remark}\label{rem:rectification}
(1)  If $S\subseteq\tfR$ is compact and $|S|<1$, then it is always possible to choose $T>0$ small enough and $L\in\N$ large enough that
$S$ admits a $(T,L)$-rectification.  In fact we can also require that for all $\epsilon>0$ sufficiently small,
$$\frac{|\Gamma|}{L} = \sum_{(q,m)\in\Gamma}|R_{q,m}| < |S|(1+\epsilon)<1.$$

\noindent (2) Under certain mild regularity assumptions on a domain $S$, we can explicitly estimate $T$ and $L$ that work.
Specifically,  $L\in\N$ can be chosen so that all such domains have a $(\sqrt{L},L)$-rectification.
\end{remark}

\begin{theorem}\label{thm:rectification2}
Fix $A,B,\epsilon,U>0$, $N\in  \N$, $0<\sigma\le 1$.  Suppose that
$S\subseteq [-A,A]{\times}[-B,B]$ and there exist $N$ Jordan curves $C_i$ such that
\begin{enumerate}
\item $S$ is contained in the interior sets of the Jordan curves,
\item the sum of areas of the interior sets is less than $\sigma-\epsilon$, and
\item the sum of lengths of the Jordan curves is bounded by $U$.
\end{enumerate}
Then for every $L$ satisfying $A,B\leq (L-1)/2$ and $4( U/\sqrt{L}+N/L)\leq \epsilon$, the set
$S+(A,B)$ has a $(\sqrt{L},L)$-rectification with $|\Gamma|\le \sigma L$.
\end{theorem}

\subsection{Sampling and reconstructing operators}
One of the contributions of this paper is to give  explicit reconstruction formulas for the impulse response of the channel operator from
the operator's response to the identifier.  Such formulas illustrate a connection between operator identification and classical sampling theory.

\subsubsection{Operators with rectangular spreading domains}

We begin by recalling a result from \cite{Pfa10}. It is a special case of Theorem~\ref{thm:reconstructionquasiperiodic} below, and is the simplest example on how Shannon's sampling theorem can be extended to apply to operators.

\begin{theorem}\label{thm:main-simple}
For $H\in OPW^2(S)$, $S\subseteq[0, T) {\times}[-\Omega/2,\Omega/2)$ compact and  $T\Omega\leq 1$,
\begin{align}
h(x,t) &= e^{-\pi i t/T}\,\sum_{n\in\Z} \biggl[\big(H\sum_{k\in\Z}\delta_{kT}\big)(t+nT) \nonumber \\
       & \hskip.5in {\times}\frac{\sin(\frac{\pi}{T}((x-t)-nT))}{\pi ((x-t)-nT)}\biggr]\CHI_{[0,T]}(t)
\label{eqn:operatorreconstruction-simple}.
\end{align}
where the sum converges in $L^{2}(\tfR)$ and for each $t$, uniformly in $x$.
\end{theorem}

\subsubsection{Non-rectangular, rectifiable spreading domains}

The following theorem gives a reconstruction formula for operators in $OPW^2(S)$ when $S$ has a rectification
in the sense of Defintion~\ref{def:rectification}.

\begin{theorem}\label{thm:reconstructionquasiperiodic}
Suppose that $S\subseteq\tfR$ and that for some $(t_0,\nu_0)$, $S-(t_0,\nu_0)$ admits a $(T,L)$-rectification,
and let $\Omega=1/(TL)$.
Then $OPW^2(S)$ can be identified by regular operator sampling,
and there exist period-$L$
sequences $b_{(q,m)} = (b_{(q,m),k})$ and functions $\Phi_{(q,m)}(t,\nu)$ for $(q,m)\in\Gamma$, such that
\begin{align}
h(x,t) &  =   e^{2\pi i (t+ t_0)\nu_0} \nonumber \\
       & \hskip-.25in \sum_k \sum_{(q,m)\in\Gamma} \big[b_{(q,m),k}\,Hg(t - (q-k)T) \nonumber \\
       & \hskip-.25in e^{-2\pi i m(q-k)/L}\,\Phi_{(q,m)}(t,x-(t+t_0)+(q-k)T)\big]. \label{eqn:reconstructionformulaperiodic}
\end{align}
where the sum converges unconditionally in $L^2(\tfR)$.
Here
$$\Phi_{(q,m)}(t,s) = \int e^{2\pi i\nu s}\,\CHI_{S_{(q,m)}}(t,\nu)\,d\nu$$
where
$$S_{(q,m)} = S\cap \bigcup_{k,\ell\in\Z} (R_{q,m}+(k/\Omega,\ell/T)).$$
\end{theorem}

\begin{remark}\label{rem:reconstructionquasiperiodic}
(1)  The coefficient sequences $b_{(q,m)}$ are defined in \eqref{eqn:leftinverse} and are the rows of a left-inverse of the
$L\times |\Gamma|$ submatrix of $G(c)$ that allows (\ref{eqn:basiclinearsystem}) to be uniquely solvable, extended to have period $L$.

\noindent (2) In light of Lemma~\ref{lem:rectification3}, it follows that for any region $S\subseteq\R^2$ for which regular operator
sampling of $OPW^2(S)$ is possible, a formula like (\ref{eqn:reconstructionformulaperiodic}) holds.  By realizing $S$ as a disjoint union
of sets $S_j$ as in (\ref{eqn:disjointunion}), each of which admits a $(T,L)$-rectification, and moreover where each $(t,\nu)\in S_j$
corresponds to the same reduced linear system in \eqref{eqn:basiclinearsystem}, we can write
$$\eta(t,\nu) = \sum_{j=1}^N \eta(t,\nu)\chi_{S_j}(t,\nu) = \sum_{j=1}^N \eta_j(t,\nu)$$
and by (\ref{eqn:symbolrelations})
$$h(x,t) = \sum_{j=1}^N \int \eta_j(t,\nu)\, e^{2\pi i \nu (x-t)}\, d\nu = \sum_{j=1}^N h_j(x,t).$$
For each $j$, we can take $t_0=\nu_0=0$ in \eqref{eqn:reconstructionformulaperiodic} and obtain
\begin{align*}
h_j(x,t) &  =   \sum_k \sum_{(q,m)\in\Gamma_j} \big[b^j_{(q,m),k}\,Hg(t - (q-k)T) \\
       & \hskip-.25in e^{-2\pi i m(q-k)/L}\,\Phi^j_{(q,m)}(t,x-t+(q-k)T)\big]
\end{align*}
where $\Gamma_j$ indexes the active boxes in the $(T,L)$-rectification of $S_j$,
$$\Phi^j_{(q,m)}(t,s) = \int e^{2\pi i\nu s}\,\CHI_{S^j_{(q,m)}}(t,\nu)\,d\nu$$
and
$$S^j_{(q,m)} = S_j\cap \bigcup_{k,\ell\in\Z} (R_{q,m}+(k/\Omega,\ell/T)).$$
Setting $b^j_{(q,m)}=0$ if $(q,m)\notin\Gamma_j$,
\begin{align}
h(x,t) &  =   \sum_{j=1}^N h_j(x,t) \nonumber \\
       &  =   \sum_{j=1}^N \sum_k \sum_{q,m=0}^{L-1} \big[b^j_{(q,m),k}\,Hg(t - (q-k)T) \nonumber\\
       & \hskip-.25in e^{-2\pi i m(q-k)/L}\,\Phi^j_{(q,m)}(t,x-t+(q-k)T)\big] \nonumber \\
       &  = \sum_k \sum_{q,m=0}^{L-1} Hg(t - (q-k)T) \nonumber \\
       & \hskip.5in \widetilde{\Phi}_{(q,m),k}(t,x-t+(q-k)T) \label{eqn:reconstructionformulaperiodic2}
\end{align}
where
$$\widetilde{\Phi}_{(q,m),k} = \sum_{j=1}^N b^j_{(q,m),k}\,e^{-2\pi i m(q-k)/L}\,\Phi^j_{(q,m)}.$$
\end{remark}


\subsubsection{Smooth reconstruction functions in the ``oversampled'' case}

Note that Theorem~\ref{thm:main-simple}, and Theorem~\ref{thm:reconstructionquasiperiodic} both involve the use of sharp cut-off
functions in the definition of the reconstruction functions $\Phi_{(q,m)}(t,s)$.  The passage to smooth cut-off and hence reconstruction
functions is enabled by the assumption that $S$ is compact with $\abs{S}<1$.  This allows for faster decay of the reconstruction functions,
and for the validity and convergence of the reconstruction sums in more general function spaces.  These matters have been studied extensively in \cite{Pfa10}.
Specifically, we have the following generalization of Theorem~\ref{thm:reconstructionquasiperiodic}.

\begin{theorem}\label{thm:reconstruction}
Suppose that $S\subseteq \R^2$, $\abs{S}<1$, is compact.  Then there exist $T>0$, $L\in\N$, $(t_0,\nu_0)$, and a period-$L$ sequence
$c=(c_n)$ such that $g=\sum_n c_n\,\delta_{nT}$ identifies $OPW^{2}(S)$.
Moreover, there exist period-$L$ sequences $b_{(q,m)}$, $(q,m)\in\Gamma$ such that
\begin{align}
h(x,t)  &=  e^{2\pi i (t+ t_0)\nu_0} \nonumber \\
        & \hskip-.25in \sum_k\sum_{(q,m)\in\Gamma} \big[b_{(q,m),k}\,Hg(t - (q-k)T) \nonumber  \\
        & \hskip-.25in e^{2\pi i m(x-t)/LT}\,\phi(x-(t+t_0)+(q-k)T)\, r(t-q T)\big] \label{eqn:reconstructionformula}
\end{align}
where  $r,\phi\in {\cal S}(\R)$ satisfy
\begin{equation}\label{eq:r_phi_2}
\sum_{k\in\Z} r(t + kT) = 1 = \sum_{n\in\Z} \widehat{\phi}(\gamma + n/LT),
\end{equation}
where $r(t)\widehat{\phi}(\gamma)$ is supported in a neighborhood of $[0,T]{\times}[0,1/LT]$, and where the
sum in \eqref{eqn:reconstructionformula} converges unconditionally in $L^{2}$ and for each $t$ uniformly in $x$.
\end{theorem}

Equation (\ref{eqn:reconstructionformula}) is a direct generalization of (\ref{eqn:operatorreconstruction-simple}) under the assumption that
$r(t)=\CHI_{[0,T]}(t)$ and $\widehat{\varphi}(\gamma) = \CHI_{[0,\Omega]}(\gamma)$.

\subsubsection{Rectification by parallelograms}

It can be advantageous to consider $S$ to be a subset of a fundamental domain of a general lattice
$A\Z^2$ where $A= \big(\begin{smallmatrix} a_{11} & a_{12} \\ a_{21} & a_{22} \end{smallmatrix}\big)$.
Our next theorem relies on basic insights on the role of symplectic geometry in time-frequency and
generalizes Theorem~\ref{thm:reconstructionquasiperiodic}.  For simplicity, we restrict our attention to lower triangular matrices
$\big(\begin{smallmatrix} a_{11} & 0 \\ a_{21} & a_{22} \end{smallmatrix}\big)$.
In Section~\ref{sec:reconstructionquasiperiodicsymplectic} we discuss the general case in detail and compute the quite involved
resulting reconstruction formulas (\ref{eqn:reconsymplecticgeneral})--(\ref{eqn:reconsymplecticgeneral03}),
(\ref{eqn:reconsymplecticgeneral04}), (\ref{eqn:reconsymplecticgeneral05}).

\begin{theorem}\label{thm:reconstructionquasiperiodicsymplectic}
Let $S\subseteq\tfR$, $\abs{S}\le 1$, and assume that with
$A= \big(\begin{smallmatrix}
 T & 0 \\ a & \Omega
\end{smallmatrix}\big)$, $\det A=T\Omega=1/L$, for some $\nu_0\in\R$, $S+(0,\nu_0)$ is contained in a fundamental
domain of the lattice $LA \Z^2$, and that
$\big(\begin{smallmatrix} T & 0 \\ 0 & 1/LT \end{smallmatrix}\big)A^{-1}(S+(0,\nu_0))$
admits a $(T,L)$-rectification, that is, if $P_{q,m} = A \big( [0,1]^2 +(q,m)^T\big)$, $q,\,m\in\Z$, then
\begin{equation}\label{eqn:rectificationofperiodizedsupportsymplectic}
LA \big([0,1]^2\big)\cap\bigcup_{k,\ell\in\Z} S+(0,\nu_0)+ LA  (k,\ell)^T \subseteq \bigcup_{(q,m)\in\Gamma} P_{q,m}.
\end{equation}
Then $OPW^2(S)$ can be identified by operator sampling. Namely, with the period-$L$ sequences $c=(c_n)$ and
$b_{(q,m)}$ from Theorem~\ref{thm:reconstructionquasiperiodic}, and functions
$$\Phi_{(q,m)}(t,s) = \int e^{2\pi i\nu s}\,\CHI_{S_{(q,m)}}(t,\nu)\,d\nu,$$
$$S_{(q,m)} = S\cap \bigcup_{k,\ell\in\Z} (P_{q,m}+LA  (k,\ell)^T),$$
\begin{align}\label{eqn:reconstructionperiodizedsupportsymplectic}
h(x,t) & =
		e^{-\pi i a t^2/T}\,
		\sum_k \sum_{(q,m)\in\Gamma} \big[b_{(q,m),k}\,
		e^{-\pi i  a T(t/T - (q-k))^2 } \nonumber \\
	   & \hskip-.25in	H g(t-  (q-k)T)\,\Phi_{(q,m)}(t , \, x  -(q-k)T) \,
	  	e^{2\pi i (q-k)at}\big].
\end{align}
Here the identifier $g=  \sum c_n e^{\pi i Ta n^2}\delta_{nT}$  and
the reconstruction sum converges unconditionally in $L^2(\tfR)$.
If the product $Ta$ is rational, say $Ta/2=p/q$ in lowest terms,
then $(c_n e^{\pi i Ta n^2})_n$ is periodic with period being the least common multiple of $q$ and $L$.
In particular, if $LTa/2$ is an integer, then the period is $L$ as well.
\end{theorem}

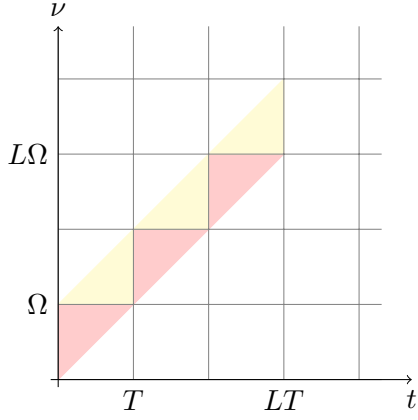
\begin{figure}
  \centering
\begin{tikzpicture}[scale=0.5]
\def\L{9}
\def\Lminusone{8}

\filldraw [fill=red!20, draw=red!20] (0,0) -- (2,2) -- (0,2) -- (0,0);
\filldraw [fill=red!20, draw=red!20] (2,2) -- (4,4) -- (2,4) -- (2,2);
\filldraw [fill=red!20, draw=red!20] (4,4) -- (6,6) -- (4,6) -- (4,4);

\filldraw [fill=yellow!20, draw=yellow!20] (0,2) -- (2,2) -- (2,4) -- (0,2);
\filldraw [fill=yellow!20, draw=yellow!20] (2,4) -- (4,4) -- (4,6) -- (2,4);
\filldraw [fill=yellow!20, draw=yellow!20] (4,6) -- (6,6) -- (6,8) -- (4,6);

\draw[style=help lines, step={(2,2)}] (0,0) grid (\L-0.4, \L+0.4);
\draw[->] (-0.2,0) -- (\L+0.4,0) node[below] {$t$};
\draw[->] (0,-0.2) -- (0,\L+0.4) node[above] {$\nu$};
\node[below] at (2,0) {$T$};
\node[left] at (0,2) {$\Omega$};
\node[below] at (6,0) {$LT$};
\node[left] at (0,6) {$L\Omega$};
\end{tikzpicture}
  \caption{The space $OPW^2(S)$ where $S$ is the union of the red and the yellow sets is identifiable with reconstruction formula
  (\ref{eqn:reconstructionperiodizedsupportsymplectic}) with a periodically weighted delta train of period $6$.
  Alternatively, we can identify $OPW^2(S)$ using a periodically weighted delta train of period $3$, but we have to solve $2$ linear systems
  or equivalently invert $2$ $3\times 3$ matrices.  In this case reconstruction is given by (\ref{eqn:reconstructionformulaperiodic2}).
  See Example~\ref{exa:rationalparallelogram} for details.}
  \label{fig:rectification3}
\end{figure}

\begin{example}\label{exa:rationalparallelogram} \rm
(1)  Figure \ref{fig:rectification3} illustrates Theorem~\ref{thm:reconstructionquasiperiodicsymplectic}.
In this case, $S$ is the union of the red and yellow triangles and hence is a parallelogram of area $1$
and $\displaystyle{A=\big(\begin{smallmatrix} T & 0 \\ \Omega & \Omega \end{smallmatrix}\big)}$ with $T\Omega=1/L=1/3$.
Theorem~\ref{thm:reconstructionquasiperiodicsymplectic} says that $OPW^2(S)$ can be identified by a periodically weighted
delta train of period $2L=6$.  However, since
$$\big(\begin{smallmatrix} T & 0 \\ 0 & \frac{1}{LT} \end{smallmatrix}\big)\,A^{-1}\,S
  = \big(\begin{smallmatrix} 1 & 0 \\ -\frac{\Omega}{T} & 1 \end{smallmatrix}\big)\,S
  =  [0,LT]\times[0,\Omega]$$
admits a $(T,L)$-rectification with $L=3$, recovery of the spreading function would only require solving a single
$3\times 3$ linear system or, equivalently, finding the three period-$3$ sequences $b_{(q,m)}$ in (\ref{eqn:reconstructionperiodizedsupportsymplectic})
would require inverting a single $3\times 3$ matrix.

\smallskip\noindent (2)  Alternatively, by considering the red and yellow regions separately as in Remark~\ref{rem:reconstructionquasiperiodic}(2),
$OPW^2(S)$ can be identified by a periodically weighted delta train of period $3$.  However, recovery of $\eta(t,\nu)$ requires the solution of
two $3\times 3$ linear systems and finding the coefficients $b^j_{(q,m)}$ in \eqref{eqn:reconstructionformulaperiodic2}
requires inverting two $3\times 3$ matrices.
\end{example}

\begin{figure}
  \centering
\begin{tikzpicture}[scale=0.5]
\def\L{9}
\def\Lminusone{8}
\filldraw [fill=blue!20, draw=blue!20] (0,0) -- (4,2.828) -- (8,6.657)-- (4,3.828) -- (0,0);
\draw[style=help lines, step={(2,2)}] (0,0) grid (\L-0.4, \L+0.4);
\draw[->] (-0.2,0) -- (\L+0.4,0) node[below] {$t$};
\draw[->] (0,-0.2) -- (0,\L+0.4) node[above] {$\nu$};
\node[below] at (2,0) {$1$};
\node[below] at (4,0) {$2$};
\node[below] at (6,0) {$3$};
\node[below] at (8,0) {$4$};
\node[left] at (0,2) {$1$};
\node[left] at (0,4) {$2$};
\node[left] at (0,6) {$3$};
\node[left] at (0,8) {$4$};
\node[below] at (4.5,-1) {(a)};
\end{tikzpicture}
\begin{tikzpicture}[scale=0.5]
\def\L{9}
\def\Lminusone{8}
\filldraw [fill=blue!20, draw=blue!20] (0,0) -- (2,1.414) -- (2,1.914) -- (0,0);
\filldraw [fill=blue!20, draw=blue!20] (0,1.414) -- (2,2.828) -- (2,3.828) -- (0,1.914);
\filldraw [fill=blue!20, draw=blue!20] (0,2.828) --(2,4.742 ) -- (2,5.242 ) -- (0,3.828) ;
\filldraw [fill=blue!20, draw=blue!20] (0,4.742 )  -- (2,6.657) -- (0,5.242 );

\draw[style=help lines, step={(2,2)}] (0,0) grid (2, \L+0.4);
\draw[-] (-0,0) -- (2.,0) node[below] {$ $};
\draw[->] (0,-0.2) -- (0,\L+0.4) node[above] {$\nu$};
\node[below] at (2,0) {$1$};
\node[below] at (1,-1) {(b)};
\node[left] at (0,2) {$1$};
\node[left] at (0,4) {$2$};
\node[left] at (0,6) {$3$};
\node[left] at (0,8) {$4$};

\end{tikzpicture}
  \caption{ (a) The the operator class $OPW^2(S)$ with $S=  ( 2,\  2 \ ;\  \sqrt{2}, \ \sqrt{2} +1/2)[0,1]^2$ whose
  area equals $1$ and bandwidth equals $1/2$ is identifiable by a (non-periodically) weighted delta train with sampling density $1/2$.
  It is not identifiable using regular operator sampling. \quad (b) $T=1$ periodization of $S$. For details, see Example~\ref{exa:non periodic}.}
\label{fig:nonperiodic}
\end{figure}
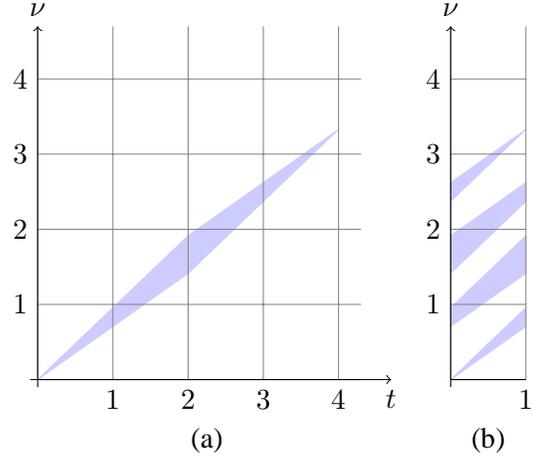

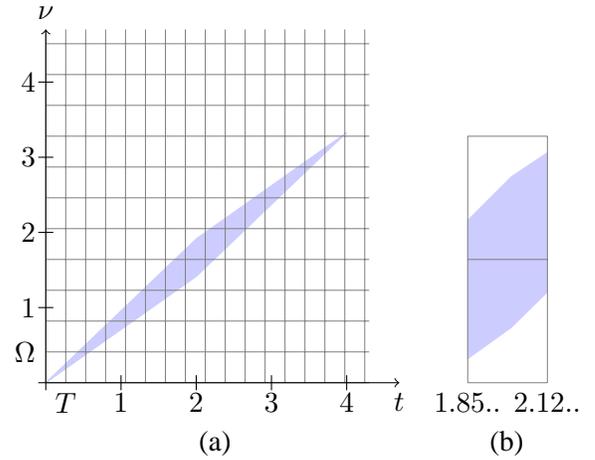
\begin{figure}
  \centering
\begin{tikzpicture}[scale=0.5]
\def\L{9}
\def\Lminusone{8}
\filldraw [fill=blue!20, draw=blue!20] (0,0) -- (4,2.828) -- (8,6.657)-- (4,3.828) -- (0,0);
\draw[style=help lines, step={(.53,.82)}] (0,0) grid (\L-0.4, \L+0.4);
\draw[->] (-0.2,0) -- (\L+0.4,0) node[below] {$t$};
\draw[->] (0,-0.2) -- (0,\L+0.4) node[above] {$\nu$};
\node[below] at (.53,0) {$T$};
\node[left] at (0,.82) {$\Omega$};
\node[below] at (2,0) {$1$};
\draw[-] (2,-0.2) -- (2,0.2);
\node[below] at (4,0) {$2$};
\draw[-] (4,-0.2) -- (4,0.2);
\node[below] at (6,0) {$3$};
\draw[-] (6,-0.2) -- (6,0.2);
\node[below] at (8,0) {$4$};
\draw[-] (8,-0.2) -- (8,0.2);
\node[left] at (0,2) {$1$};
\draw[-] (-0.2,2) -- (0.2,2);
\node[left] at (0,4) {$2$};
\draw[-] (-0.2,4) -- (0.2,4);
\node[left] at (0,6) {$3$};
\draw[-] (-0.2,6) -- (0.2,6);
\node[left] at (0,8) {$4$};
\draw[-] (-0.2,8) -- (0.2,8);
\node[below] at (4.5,-1) {(a)};

\end{tikzpicture}
\begin{tikzpicture}[scale=2]
\def\L{9}
\def\Lminusone{8}
\filldraw [fill=blue!20, draw=blue!20] (3.71-7*.53,2.62-3*.82) -- (4-7*.53,2.828-3*.82) -- (0.53,.605) -- (0.53,1.53) -- (4-7*.53,1.37) -- (0,1.08) -- (3.71-7*.53,2.62-3*.82);
\draw[style=help lines, step={(.53,.82)}] (0,0) grid (.53,2*.82);
\node[below] at (0,0) {$1.85..$};
\node[below] at (.53,0) {$2.12..$};
\node[below] at (.26,-.25) {(b)};
\end{tikzpicture}

  \caption{(a) For periodic operator sampling to succeed with $S$ having area 1, we require that the $T,\Omega$ periodization of $S$
  leads to an exact $L$ cover of the time-frequency plane. \quad (b) The central piece of the set $S$. For the significance of this set,
  see Example~\ref{exa:non periodic}.}  \label{fig:blowup}
\end{figure}

\begin{example}\label{exa:non periodic} \rm
Figures~\ref{fig:nonperiodic} and \ref{fig:blowup} illustrate a situation in which $OPW^2(S)$ can be identified by operator sampling
but not by regular operator sampling.  In this case,
$\displaystyle{A=\big(\begin{smallmatrix} 2 & 2 \\ \sqrt{2} & \sqrt{2}+1/2 \end{smallmatrix}\big)}$, $S=\displaystyle{A[0,1]^2}$
and hence $A^{-1}S = [0,1]^2$ admits a $(T,L)$-rectification with $T=L=1$.  Therefore, following the notation in the proof of
Theorem~\ref{thm:reconstructionquasiperiodicsymplectic}, $A=B$, $c_n=1$ for all $n$, $L'=2$, and $c'_n = 1-e^{\pi in}$.  By
equation (\ref{eqn:nonperiodicidentifier}), $OPW^2(S)$ can be identified by
$$g = \mu(B)\widetilde{g} = \frac{1}{\sqrt{2}}\sum (1-e^{\pi in})e^{\pi in^2 \sqrt{2}/2}\,\delta_{n},$$
a delta train with non-periodic weights.

Note that $B(S)=1/2$, and that since $1-e^{\pi in}=0$ when $n$ is even, the sampling density
of the identifier $g$ is also $1/2$.  Therefore, by Theorem~\ref{thm:necessarysamplingrate}, this identifier achieves the minimal
sampling rate for this region.

Next we observe that this region cannot be identified by regular operator sampling for any value of $T$ or $L$.
Since $|S|=1$, by Remark~\ref{rem:characterization}(2), the $T\Z\times\Omega\Z$-periodization of $S$ must be an exact $L$-cover.
In other words, the inequality in (\ref{eqn:characterization2}) must be an equality.  It can be shown, however, that for any
value of $T$ and $L$, this is not possible.  Details of the argument can be found in Section~\ref{sec:parallelograms}.
\end{example}

\subsection{Operator Sampling as a Generalization of Classical Sampling.}\label{section:classicalsampling}

By generalizing the setting to other function spaces, we can more precisely illustrate the connection between
operator sampling and the classical sampling theorem of Shannon, Whittaker, and Kotelnikov among others, and also the connection with
the well-known fact that a time-invariant operator can be identified by its impulse response.

\begin{definition}\label{generaloperatorpaleywienerspaces}
We define the operator Paley-Wiener spaces $OPW^{\infty,2}(S)$ and $OPW^{2,\infty}(S)$ by
\begin{align*}
OPW^{\infty,2}(S) & =  \{H\in \mathcal L (L^2(\R), L^2(\R)) \colon \\
      & \hskip.25in  \supp\eta_H\subseteq S,\,\norm{\sigma_H}_{L^{\infty,2}}<\infty\}
\end{align*}
where
$$\norm{\sigma_H}_{L^{\infty,2}} = \Bigl\|\int\abs{\sigma_H (\cdot,\xi)}^2 d\xi\Bigr\|^{1/2}_\infty$$
and
\begin{align*}
OPW^{2,\infty}(S) & =  \{H\in \mathcal L (L^2(\R), L^2(\R)) \colon  \\
      & \hskip.25in \supp\eta_H\subseteq S,\,\norm{\sigma_H}_{L^{2,\infty}}<\infty\}
\end{align*}
where
$$\norm{\sigma_H}_{L^{2,\infty}} = \Big(\int \norm{\sigma_H (x,\cdot)}^2_\infty dx\Big)^{1/2}$$
(\cite{Pfa10}, Theorem~4.2).
$OPW^{p,q}(S)$ is a Banach space with respect to the norm $\norm{H}_{OPW^{p,q}} = \norm{\sigma_H}_{L^{pq}}$.
\end{definition}

Note that convolution
with a compactly supported kernel whose Fourier transform is in $L^2$ is an operator in $OPW^{\infty,2}$
and multiplication by a bandlimited function in $L^2$ is an operator in $OPW^{2,\infty}$.

\subsubsection{Identification of convolution operators}\label{sec:classicalsampling1}

First, take $H$ to be ordinary convolution by $h_H(t)$, that is, $h_H(x,t) = h_H(t)$.  In this case $H$
can be identified in principle by $g=\delta_0$, the unit impulse at the origin, since $Hg(x) = h_H(x)$.
That is, $\Lambda = \set{0}$ is a sampling set for the class of convolution operators.
Translating this into our operator sampling formalism results in something slightly different.

Assume that $h$ is supported in the interval $[0,T']$, $\widehat h \in L^2$ and that $T>T'$, and $\Omega>0$
are chosen so that $\Omega T<1$.  In this case, $\eta_H(t,\nu) = h(t)\,\delta_0(\nu)$ and
$\sigma_H(x,\xi) = \widehat{h}(\xi)$.  Therefore $\sigma_H\in L^{\infty,2}$
and $H\in OPW^{\infty,2}([0,T']{\times}[-\Omega/2,\Omega/2])$.

Applying Theorem~\ref{thm:reconstruction} to this situation, note that
if $g=\sum_n\delta_{nT}$ then $Hg$ is simply the $T$--periodized impulse response $h(t)$, and
it follows from the theorem (or by direct calculation) that with $r,\,\varphi\in{\cal S}(\R)$,
$r(t)=1$ on $[0,T']$ and vanishing outside an interval of length $T$
containing $[0,T']$, and with $\widehat{\varphi}(0)=1$ and $\supp\widehat{\varphi}\subseteq [-\Omega/2,\Omega/2]$,
\begin{align*}
 & r(t)\sum_{k\in\Z} (Hg)(t+kT)\varphi(x-t-kT) \\
 &\hskip.1in = \sum_{k\in\Z}  \sum_{n\in\Z}r(t)\,h(t+kT-nT)\,\varphi(x-t-kT)\\
 &\hskip.1in = \sum_{k\in\Z}   h(t)\,\varphi(x-t-kT)=  h(t).
\end{align*}
Here we have used the fact that $r(t)=1$ on $[0,T']$ and vanishes outside a neighborhood of $[0,T']$ and that
$\sum_k\varphi(x-t-kT)=1$ by the Poisson Summation Formula and in consideration of the support
constraints on $\widehat{\varphi}$.  Indeed the theorem says that the sum $\sum_k\varphi(x-t-kT)$
converges to $1$ in the $L^\infty$ norm and in particular uniformly on compact sets.

\subsubsection{Identification of multiplication operators (Classical Sampling)}\label{sec:classicalsampling2}

To compare Theorem~\ref{thm:reconstruction} with the classical sampling theorem,
take $H$ to be multiplication by some fixed function $m\in L^2$ with
$\supp\widehat{m}\subseteq [-\Omega/2,\Omega/2]$ then $\eta_H(t,\nu) = \delta_0(t)\widehat{m}(\nu)$,
$h(t,x) = \delta_0(t)\,m(x-t)$, and $\sigma_H(x,\xi) = m(x)$.  Let $\Omega'>\Omega$ and $T>0$ be such that $\Omega' T<1$.
Then $\sigma_H\in L^{2,\infty}$ and $H\in OPW^{2,\infty}([-T/2,T/2]\times[-\Omega/2,\Omega/2])$.

Choose $r,\,\varphi\in{\cal S}(\R)$ such that $\supp r\subseteq [-T/2,T/2]$, $r(0)=1$,
$\supp\widehat{\varphi}\subseteq [-\Omega'/2,\Omega'/2]$, and $\widehat{\varphi}(\nu) = 1$ on $[-\Omega/2,\Omega/2]$.
If $g=\sum_n\delta_{nT}$, then $Hg = \sum_n m(nT)\,\delta_{nT}$, and
it follows from Theorem~\ref{thm:reconstruction} (and by direct calculation) that
\begin{align*}
&  \delta_0(t)\,m(x-t) \\
&  \hskip.25in =    r(t)\sum_{k\in\Z} (Hg)(t+kT)\varphi(x-t-kT) \\
&  \hskip.25in =    r(t)\sum_{k\in\Z} \sum_{n\in\Z} m(nT)\,\delta_{(n-k)T}(t)\varphi(x-t-kT) \\
&  \hskip.25in =    \sum_{n\in\Z} m(nT)\,\varphi(x-nT)
\end{align*}
by support considerations on the function $r(t)$.  Therefore we have the summation formula
$$m(x) = \sum_{n\in\Z} m(nT)\,\varphi(x-nT)$$
where the sum converges unconditionally in $L^2$. This recovers the classical sampling formula when sampling above the Nyquist rate.

\subsection{Sufficient conditions on the sampling rate in operator sampling}\label{section:samplingrate}

As was observed earlier, a natural measure of the sampling rate in operator sampling is the quantity $D(\Lambda)$ (Definition~\ref{def:samplingset}),
which in the case of regular operator sampling is $\norm{c}_0/(TL)$.  A necessary condition on the sampling rate
in operator sampling was give in terms of the bandwidth of a channel (Theorem~\ref{thm:necessarysamplingrate}).  The goal of this subsection
is to investigate sufficient conditions on the sampling rate in regular operator sampling that guarantee identifiability.

In the classical sampling theory of functions,
the sampling rate must exceed the reciprocal of the area of the bandlimiting set; and regardless of the measure of the bandlimiting set,
a (possibly high density) sampling set always exists.  As mentioned above (Theorem~\ref{thm:OPWidentifiable}),
operator sampling of $OPW^2(S)$ is only possible if the measure of
$S$ satisfies $\abs{S}\leq 1$, and necessary sampling rates in operator sampling depend on the geometry of $S$.

The main result in this paper relevant to finding a sufficient condition on the sampling rate for identification of $OPW^2(S)$ is the following.

\begin{theorem}\label{thm:sufficientsamplingrate}
Let $S\subseteq\tfR$ be compact, $|S|<1$, $\epsilon>0$, and suppose that $S$ has a $(T,N)$-rectification
satisfying $(\abs{\Gamma}+2)/N<\abs{S}(1+\epsilon)<1$.
Then for every sufficiently large $L\in\N$,
$OPW^2(S)$ can be identified via regular operator sampling by an identifier $g=\sum_n c_n\,\delta_{nT}$, where
$c=(c_n)$ is a period-$L$ sequence satisfying
$$\frac{\norm{c}_0}{L} < \abs{S}(1+\epsilon).$$
Moreover, if $L$ is prime, then $c$ can be chosen with $c_j=0$ if $\norm{c}_0\le j<L$, that is, such that $c$
is supported on its first $\norm{c}_0$ indices.
\end{theorem}

\begin{remark}\label{rem:sufficientsamplingrate}
(1)  Note that once an appropriate $(T,N)$-rectification of $S$ is found,
the parameter $T$ associated to that rectification is fixed.  Subsequently, a periodic weighting sequence can be found
for the delta train $\sum_n c_n \delta_{nT}$ whose relative support is bounded essentially by the area of $S$.
Moreover, if $L$ is prime, and $c$ is supported on $\set{0,\,1,\,\dots,\,\norm{c}_0-1}$, then this represents
a {\em bunched operator sampling} that can allow
for the efficient identification of the channel in the following way.

If the area of $S$ is small, and if $K$ represents the ``memory'' of the channel (that is, for each $\nu$, $\eta(t,\nu)$ is supported
in the interval $[0,K]$), then the response of the channel to the delta train $\sum_n c_n\delta_{nT}$ is supported on the set
$$\bigcup_{j\in\Z} \big([0,T\norm{c}_0+K]+jLT\big)$$
and hence vanishes on the set
$$\bigcup_{j\in\Z} \big([T\norm{c}_0+K, LT]+jLT\big).$$
The ``dead time'' represented by this set can be used for other purposes.
Note also that $|LT-(T\norm{c}_0+K)| \ge LT(1-|S|(1+\epsilon)-K/(LT))$ so that the length of the dead time within each period
of the channel response increases with $L$.

\noindent (2) Another interpretation of this result is that the sparsity of the matrix $G(c)$ in the linear system (\ref{eqn:basiclinearsystem})
can be controlled by the area of the spreading support.  In this case, $\norm{c}_0/L$ gives the fraction of nonvanishing entries in each
column of $G(c)$.  Hence $S$ with small support guarantees that $G(c)$ can be chosen to be sparse.
\end{remark}

\begin{remark}\label{rem:capacity}
The ``dead time'' referred to above can be thought of as a measure of the capacity of the unknown, bandlimited
channel in the sense that only during this time
can data be sent over the channel.  With this notion of capacity, the above discussion says that the capacity of a time-varying channel decays linearly
with the area of its spreading support.
\end{remark}

\subsection{Sampling and reconstruction of operators with small, but unknown support}\label{sec:unknownsupport}

Just as in classical sampling, operator sampling requires full knowledge of the bandlimitation we expect an operator to have,
that is, the reconstruction formulas for $OPW^2(S)$ depend on knowing the region $S$.
However, in some applications $S$ may not be known precisely, but only some information on its size,
geometry and location is given by physical considerations. In this section
we address the question whether such operators can be sampled and reconstructed in a stable matter.

\begin{theorem}\label{thm:unknownDomain}
For $A,B,U,\epsilon,\sigma>0$ and $N\in  \N$, let ${\cal H} (A,B,U,N,\epsilon,\sigma)$ contain all operators such that
$\supp \, \mathcal F_s\sigma_H =\supp \eta_H \subseteq [-A,A]{\times}[-B,B]$ satisfies the hypothesis of Theorem~\ref{thm:rectification2}
with $\sigma\le 1/2$.
Then there exists $L\in\N$ and an $L$-periodic sequence $(c_n)$ such that
$g=\sum_n c_n \delta_{n/\sqrt{L}}$ identifies ${\cal H} (A,B,U,N,\epsilon,\sigma)$.
\end{theorem}

The reconstruction of an operator $H \in {\cal H} (A,B,U,N,\epsilon,\sigma)$ is then carried out as follows.
First choose $L$ as in Theorem~\ref{thm:rectification2} and let $R_H$ denote the {\em rectified support} of $H$, that is, the union of
$(1/\sqrt{L})\times(1/\sqrt{L})$ boxes that cover $\supp\eta_H$ having area not greater than $1/2$.
Under this assumption, we determine $R_H$.
In the final step, we apply the operator reconstruction formula developed in Theorem~\ref{thm:reconstructionquasiperiodic} to $OPW^2(R_H)$.

To determine the rectified support of $\eta_H$ with $H \in {\cal H} (A,B,U,N,\epsilon,\sigma)$, we will apply ideas from compressed sensing.
Indeed, Lemma~\ref{lem:matrixequationquasiperiodic} below, shows that from $H\sum_n c_n \delta_{n\sqrt{L}}$, we can compute a length $L$ vector
${\bf y}(t,\nu)$ with ${\bf y}(t,\nu)=G(c){\bf x}(t,\nu)$ and where the unknown discrete support of the
length $L^2$ vector ${\bf x}(t,\nu)$ encodes the support
of the bivariate function $\eta_H(t,\nu)$.  In fact, recovering the vector ${\bf x}(t,\nu)$ for a single point $(t,\nu)$  provides us with the support
structure of $\eta_H$. Note that the conditions given above imply that ${\bf x}(t,\nu)$ has at most $L/2$ nonzero components.

The full-Spark matrix $G(c)$ plays the role of a
{\em measurement matrix} and has the  ability to recover any $L/2$-sparse vector ${\bf x}(t,\nu)$  \cite{KPR08,LPW05}.  But finding an $L/2$-sparse
vector requires consideration of every support structure out of $\choo{L^2}{L/2}$ possible ones, which is hardly possible for
$L$ not being of the order $2,3,5$. If we know that far fewer than $L/2$ cells are active, then we can try to apply
compressed sensing algorithms such as Basis Pursuit or Orthogonal Matching Pursuit to recover ${\bf x}$ from ${\bf y}=G(c){\bf x}$.
See \cite{BH11}, \cite{BH13}, \cite{FB96}, and \cite{B08} for descriptions of the recovery algorithms.


In light of Theorem~\ref{thm:characterization} we can extend Theorem~\ref{thm:unknownDomain} in a different direction
and obtain a large class of operators that can be identified via regular operator sampling without knowledge of the support set.
This class is larger than the class of area $\le 1/2$ considered in \cite{BH11, BH13}.

\begin{theorem}\label{thm:characterizationunknownsupport}
Let $T>0$, $L\in\N$ and $c\in\C^L$ be chosen so that $G(c)$ has full Spark, and let $g=\sum_{n\in\Z} c_n \delta_{nT}$.
For $0\le\Delta\le 1$, define the operator class ${\cal H}_{T,L}(\Delta)$ to be the collection of operators $H$ in $OPW^2(\R^2)$
such that for some fixed fundamental domain $R$ of the lattice $(TL)\Z\times (1/T)\Z$ ,
$\supp\eta_H=S_H\subseteq R$ and
 \begin{equation}\label{eqn:characterization3}
 \sum_{k,\ell} \chi_{S_H+(kT,\ell/(TL))}\leq \Delta\,L\quad a.e.
 \end{equation}
 Then the following are equivalent.
\begin{enumerate}
\item[{\rm (i)}]
$\Delta < 1/2 + 1/(2L)$.
 \item[{\rm (ii)}]
 For all $H_1,\,H_2\in {\mathcal H}_{T,L}(\Delta)$, $H_1g=H_2g$ implies $H_1=H_2$
 \item [{\rm (iii)}]
${\mathcal H}_{T,L}(\Delta)$ is identifiable with identifier $g$ in the sense of
\eqref{eqn:identifyNONLINEAR}
\end{enumerate}
\end{theorem}

Boelcskei and Heckel (\cite{BH11, BH13}) have shown that, for operator classes like those in Theorem~\ref{thm:unknownDomain},
if only $L-1$ cells are active, these can be determined, and hence the operator class can be identified without knowing the spreading support.
Their analysis and derived recovery algorithms rely on the fact that by varying $(t,\nu)$ you obtain a family of equations
${\bf y}(t,\nu)=G(c){\bf x}(t,\nu)$ where the vectors ${\bf x}(t,\nu)$ have identical sparsity structure.  This allows for the recovery
of almost every operator (\cite{FB96}, \cite{B08}) in the given class.

In Theorem~\ref{thm:characterization} we give up joint sparsity, i.e., the sparsity structure of ${\bf x}(t,\nu)$ varies with $(t,\nu)$.
A compromise based on the characterization found in Lemma~\ref{lem:rectification3} that guarantees joint sparsity and allows us to use
Theorem~\ref{thm:characterization} is given by the following generalization of Theorem~3 in \cite{BH11} (cf. Theorem~3 in \cite{BH13}).
Note that the additional parameter $K$ can be chosen independently
of $T$ and $L$, i.e., choosing $K$ large does not increase the sampling rate, nor the size of the compressive sensing problem, i.e., of
the matrix $G(c)$.

\begin{theorem}\label{thm:unknowndomain2}
For $T>0$, and $L\in\N$, let $c\in\C^L$ be chosen so that $G(c)$ has full Spark.
Given $K\in\N$, define the operator class ${\cal H}_{T,L,K}\subseteq OPW(\R^2)$ by $H\in{\cal H}_{T,L,K}$
if and only if $\supp\eta_H = S_H\subseteq [0,LT]\times[0,1/T]$ satisfies
\begin{enumerate}
\item[{\rm (a)}] $\displaystyle{\sum_{k,\ell} \chi_{S_H+(kT,\ell/(TL))}\leq L-1\quad a.e.}$, and
\item[{\rm (b)}] each set $A_j$ in the partition of $[0,T]\times[0,1/(LT)]$ given in Lemma~\ref{lem:rectification3} can be written as a union
 of sets of the form $[0,T/K]\times[0,1/(KLT)]+(qT/K,m/(KLT))$, $0\le q,m<K$.
\end{enumerate}
Then almost every operator in $H\in{\cal H}_{T,L,K}$ can be identified by regular operator sampling.
\end{theorem}

Note that alternatively to choosing $K$, we could attempt to introduce joint sparsity by assuming that, for example, $\eta_H$ is smooth.

\section{Proofs of Theorems}

\subsection{Proof of Theorem~\ref{thm:necessarysamplingrate}}

Since $S$ is closed, each $t$-section $S_t$ of $S$ is closed and, hence, measurable. Therefore, $\chi_{S}(t,\cdot)$ is a
nonnegative measurable function and $\int_\R \chi_{S}(t,\nu)\,d\nu\in[0,\infty]$ is well defined for all $t\in\R$.
It suffices to show the result for $A_\infty=\big\|\int_\R \chi_{S}(\cdot,\nu)\,d\nu\big\|_\infty$
finite, the infinite case then follows from this.

Assume that $\Lambda$ is a set of sampling with $D(\Lambda)<a_\infty<A_\infty$.

Then, we can choose a set $P$ with positive measure and $\int_\R \chi_{S}(t,\nu)\,d\nu\geq a_\infty $ for all $t\in P$.
Assume without loss of generality $P\subseteq [0,1]$. For any $\epsilon$, there exist  $m_t\in PW(S_{t})$ with $\|m_t\|_{L^2}=1$
and $\|m_t|_\Lambda\|_{\ell^2}\leq \epsilon$, $t\in P$
. Define $\kappa_H(x,y)=  m_{x-y}(y)$ for $x-y\in P$, and $0$ otherwise.
Then $h_H(x,t)= \kappa_H(x,x-t)=m_t(x-t)$ and $\eta_H(t,\nu)=\widehat m_t(\nu)$ for $t\in P$, and $0$ otherwise,
so $H\in OPW^2(S)$.
Observe that $\|\sigma_H\|_{L^2}=\sqrt{|P|}$.
Note that it is easily seen that if $\sum_{\lambda\in\Lambda} c_\lambda \delta_\lambda$ identifies $OPW^2(S)$, then $(c_\lambda)$ is bounded.
Also, by hypothesis, there exists $K\in\N$ which bounds the cardinality of $\Lambda\cap [x,x+1]$ above for all $x\in\R$. We compute
\begin{align*}
  \big\| H \sum_{\lambda\in\Lambda} c_\lambda \delta_\lambda \big\|_{L^2}^2
  &=
 \int \big| \sum_{\lambda\in\Lambda} c_\lambda \kappa_H(x,\lambda)\big|^2\,dx\\
 	& \hskip-.1in =\int \big |\sum_{\lambda\in\Lambda} c_\lambda m_{x-\lambda}(\lambda)\big|^2\,dx\\
	& \hskip-.1in \leq\|(c_\lambda)\|^2_{\ell^\infty}\, \int \big |\sum_{\lambda\in\Lambda}   m_{x-\lambda}(\lambda)\big|^2\,dx\\
	& \hskip-.1in \leq\|(c_\lambda)\|^2_{\ell^\infty}\, K\,\sum_{\lambda\in\Lambda} \int     | m_{x-\lambda}(\lambda)|^2\,dx\\
	& \hskip-.1in =\|(c_\lambda)\|^2_{\ell^\infty}\, K\,\sum_{\lambda\in\Lambda} \int_\lambda^{\lambda+1}   | m_{x-\lambda}(\lambda)|^2\,dx\\
	& \hskip-.1in =\|(c_\lambda)\|^2_{\ell^\infty}\, K\, \int_0^1 \sum_{\lambda\in\Lambda}  | m_{t}(\lambda)|^2\,dt\\
	& \hskip-.1in \leq \|(c_\lambda)\|^2_{\ell^\infty}\, K\, \int_0^1   \epsilon^2\,dt = \|(c_\lambda)\|^2_{\ell^\infty}\, K\, \epsilon^2\,.\\
\end{align*}

\subsection{Proof of Equation~\eqref{eqn:basiclinearsystem}}

\begin{definition}
  The non-normalized Zak Transform is defined for $f\in{\cal S}(\R)$, and $a>0$ by
$$\displaystyle{Z_a f(t,\nu) = \sum_{n\in\Z} f(t-an)\,e^{2\pi ia n\nu}}.$$
\end{definition}

$Z_af(t,\nu)$ satisfies the quasi-periodicity relations
$$\displaystyle{Z_af(t+a,\nu) = e^{2\pi ia\nu}\,Z_af(t,\nu)}$$
and
$$\displaystyle{Z_af(t,\nu+1/a) = Z_af(t,\nu)}.$$  $\sqrt{a}\,Z_a$ can be extended to a unitary operator
from $L^2(\R)$ onto $L^2([0,a]{\times}[0,1/a])$.

The following Lemma connects the output $Hg(x)$ where $g$ is a delta-train, to
the spreading function $\eta_H(t,\nu)$.

\begin{lemma}\label{lem:deltatrain}
Let $a>0$ be given and let $g = \sum_{n}\delta_{na}$.  Then for all $(t,\nu)\in\R^2$,
\begin{align*}
(Z_a\circ H)g(t,\nu) & \\
  & \hskip-.5in = a^{-1}\,\sum_{k}\sum_{m} \eta_H(t+ak,\nu+m/a)\,e^{-2\pi i\nu ka},
\end{align*}
where $\eta_H$ is the spreading function of the operator $H$.
\end{lemma}

\begin{IEEEproof}
It can be verified by direct calculation that if $g = \sum_{n}\delta_{na}$ then
$\ip{Hg}{f} = \ip{\eta_H}{Z_a f}$ for all $f\in{\cal S}(\R)$ where the bracket on the left is the $L^2$ inner product on
$\R$ and that on the right the $L^2$ inner product on the rectangle $[0,a]{\times}[0,1/a]$.
Periodizing the integral defining the $L^2$ inner product on the left gives
\begin{align*}
\ip{\eta_H}{Z_a f} = & \int_0^{1/a}\int_0^a \sum_{k}\sum_{m} \eta_H(t+ka,\nu+m/a) \\
&  \hskip.5in  e^{-2\pi i\nu ka}\overline{Z_a f(t,\nu)}\,dt\,d\nu.
\end{align*}
Since this holds for every $f\in{\cal S}(\R)$, the result follows.
\end{IEEEproof}

\begin{lemma}\label{lem:weighteddeltatrain}
Let $T,\Omega>0$ be given such that $T\Omega = 1/L$ for some $L\in\N$, let $(c_n)$ be a period-$L$ sequence,
and define $g = \sum_n c_n\,\delta_{nT}$.  Then for $(t,\nu)\in\R\times\widehat{\R}$,
\begin{align}
(Z_{1/\Omega}\circ H)g(t,\nu) & \nonumber \\
    & \hskip-.75in = \Omega\,\sum_{q=0}^{L-1} c_{-q}\,\sum_{k}\sum_{m} \eta_H(t+k/\Omega+qT,\nu+m\Omega) \nonumber \\
    & \hskip.75in e^{-2\pi i(\nu+m\Omega)qT}\,e^{-2\pi i\nu k/\Omega}. \label{eqn:weighteddeltatrain}
\end{align}
\end{lemma}

\begin{IEEEproof}
Note first that letting $j=nL-q$, $0\le q\le L-1$, $n\in\Z$,
\begin{align*}
g &  =    \sum c_j\,\delta_{nT} =    \sum_{q=0}^{L-1} \sum_{n\in\Z} c_{nL-q}\,\delta_{nLT - qT} \\
   &  =    \sum_{q=0}^{L-1}  c_{-q} T_{-q/L\Omega}\,\bigg(\sum_{n\in\Z}\,\delta_{n/\Omega}\bigg).
\end{align*}
For $\alpha\in\R$, the spreading function of $H\circ T_\alpha$ is $\eta_H(t-\alpha,\nu)\,e^{2\pi i\nu\alpha}$
and hence
\begin{align*}
&       (Z_{1/\Omega}\circ H)\bigg(\sum c_j\,\delta_{nT}\bigg)(t,\nu) \\
& \hskip.1in  =  \sum_{q=0}^{L-1}  c_{-q} (Z_{1/\Omega}\circ H \circ T_{-q/L\Omega})\biggl(\sum_{n\in\Z}\,\delta_{n/\Omega}\biggr)(t,\nu).
\end{align*}
Lemma~\ref{lem:deltatrain} yields the result.
\end{IEEEproof}

Changing summation indices in (\ref{eqn:weighteddeltatrain}) by $m=nL+\ell$, $0\le\ell\le L-1$, $n\in\Z$, yields the following lemma.

\begin{lemma}\label{lem:weighteddeltatrainperiodic}
Let $T,\Omega>0$ be given such that $T\Omega = 1/L$ for some $L\in\N$, let $(c_n)$ be a period-$L$ sequence.
Then with $g = \sum_n c_n\,\delta_{nT}$, and for all $(t,\nu)\in\R^2$,
\begin{align}
& (Z_{1/\Omega}\circ H)g(t,\nu) \nonumber \\
& \hskip.1in = \Omega\,\sum_{q=0}^{L-1} c_{-q}\,\sum_{m=0}^{L-1}\sum_{k} \eta^{QP}_H(t+qT,\nu+m\Omega) \nonumber \\
& \hskip1in   e^{-2\pi i\nu qT}\,e^{-2\pi i\nu mq/L} \label{eqn:weighteddeltatrainperiodic}
\end{align}
where $\eta^{QP}_H(t,\nu)$ is the {\em ($1/\Omega,\,1/T)$--quasiperiodization} of $\eta_H$ defined below.
\end{lemma}

\begin{definition}\label{def:quasiperiodization}
Given a bivariate function $f(t,\nu)$ and parameters $T,\Omega>0$, define the {\em ($1/\Omega,\,1/T)$--quasiperiodization} of $f$, denoted $f^{QP}$, by
\begin{equation}\label{eqn:quasiperiodization}
f^{QP}(t,\nu) = \sum_k\sum_\ell f(t+k/\Omega,\nu+\ell/T)\,e^{-2\pi i\nu k/\Omega}
\end{equation}
whenever the sum is defined.  Note that $f^{QP}(t,\nu+1/T) = f^{QP}(t,\nu)$ and $f^{QP}(t+1/\Omega,\nu) = e^{2\pi i\nu/\Omega}\,f^{QP}(t,\nu)$
for all $(t,\nu)\in\R^2$.
\end{definition}

\begin{lemma}\label{lem:reconstructionfromquasiperiodization}
Suppose that $\supp(f)=S$ is contained in a fundamental domain of $1/\Omega\,\Z \times 1/T\,\Z$.  Then
\begin{align}
f(t,\nu)
 & = \sum_k \sum_\ell f^{QP}(t-k/\Omega,\nu-\ell/T) \nonumber \\
 & \hskip-.25in \CHI_{[0,1/\Omega]}(t-k/\Omega)\,\CHI_{[0,1/T]}(\nu-\ell/T)\,e^{2\pi ik\nu/\Omega}\,\CHI_S(t,\nu)
 \label{eqn:reconstructionfromquasiperiodization}
\end{align}
where if $f\in L^2(\R^2)$, the sum converges in $L^2$ and uniformly on compact sets.
\end{lemma}

\begin{IEEEproof}
Under the given assumptions, the functions being summed in (\ref{eqn:reconstructionfromquasiperiodization})
have pairwise disjoint supports. Since $\abs{S}<1$, the sum converges in $L^2$ if $f\in L^2(\R^2)$.
Moreover, since on each compact set, the sum is finite, we get uniform convergence on compact sets.

To complete the proof, we show that (\ref{eqn:reconstructionfromquasiperiodization}) holds pointwise.  Since $S$ is a fundamental domain,
for$(t,\nu)\in S$ only the $(k,\ell)=(0,0)$ term survives in (\ref{eqn:quasiperiodization}).  Hence, for all $(t,\nu)$,
$$f^{QP}(t,\nu)\,\CHI_S(t,\nu) = f(t,\nu).$$
By direct calculation,
\begin{align*}
& f^{QP}(t,\nu) = \sum_k \sum_\ell f^{QP}(t-k/\Omega,\nu-\ell/T) \\
& \hskip.25in \CHI_{[0,1/\Omega]}(t-k/\Omega)\ \CHI_{[0,1/T]}(\nu-\ell/T)\,e^{2\pi ik\nu/\Omega}
\end{align*}
for each $(t,\nu)\in\tfR$.
\end{IEEEproof}

\begin{lemma}\label{lem:matrixequationquasiperiodic}
Let $T,\Omega>0$ be given such that $T\Omega = 1/L$ for some $L\in\N$, let $(c_n)$ be a period-$L$ sequence.
Then with $g = \sum_n c_n\,\delta_{nT}$, $(t,\nu)\in\R^2$, and $p=0,\,1,\,\dots,\,L{-}1$,
\begin{align}\label{eqn:matrixequationquasiperiodic}
&       e^{-2\pi i\nu T p}\,(Z_{1/\Omega}\circ H)g(t + T p,\nu) \nonumber\\
&  =    \Omega\,\sum_{q,\,m=0}^{L-1} (T^q\,M^m c)_p\,
  e^{-2\pi i\nu T q}\,\eta^{QP}_H(t + T q,\nu + \Omega m).
  \end{align}
\end{lemma}

\begin{IEEEproof}
By (\ref{eqn:weighteddeltatrainperiodic}),
\begin{align*}
&       (Z_{1/\Omega}\circ H)g(t+pT,\nu) \\
&  =    \Omega\,\sum_{q=0}^{L-1} c_{-q}\,\sum_{\ell=0}^{L-1}\eta^{QP}_H(t+(q+p)T,\nu+m\Omega) \\
& \hskip.5in  e^{-2\pi i\nu qT}\,e^{-2\pi i\nu mq/L}.
\end{align*}
Making the change of index $q\mapsto q-p$, rearranging terms and using the fact that $LT=1/\Omega$ yields
\begin{align*}
& (Z_{1/\Omega}\circ H)g(t+Tp,\nu) \\
&  =    \Omega\,\sum_{q=0}^{L-1}\sum_{m=0}^{L-1} c_{-(q-p)}\,e^{-2\pi i m (q-p)/L} \\
&    \hskip.5in \eta^{QP}_H(t+qT,\nu+m\Omega)\,e^{-2\pi i \nu (q-p)T}.
\end{align*}
Since $(T^q\,M^m c)_p = c_{p-q}\,e^{2\pi i m (p-q)/L}$, the result follows.
\end{IEEEproof}

Letting
\begin{equation}\label{eqn:boldZvector}
{\bf Z}_{Hg}(t,\nu)_p = (Z_{1/\Omega}\circ H)g(t+pT,\nu)\,e^{-2\pi i \nu pT}
\end{equation}
and
\begin{equation}\label{eqn:boldetavector}
\boldsymbol\eta_H(t,\nu)_{(q,m)} = \Omega\,\eta^{QP}_H(t+qT,\nu+m\Omega)\,e^{-2\pi i \nu qT}\,e^{-2\pi iqm/L},
\end{equation}
we have that
$${\bf Z}_{Hg}(t,\nu)_p = \sum_{q,m=0}^{L-1} G(c)_{p,(q,m)}\,\boldsymbol\eta_H(t,\nu)_{(q,m)}$$
which is \eqref{eqn:basiclinearsystem}.

\subsection{Proof of Theorem~\ref{thm:smallsparksampta13}}

We first recall and outline the proof of Theorem~\ref{thm:lpw05}.

Given any square submatrix of $G(c)$, call it $M$, $\det(M)$ is a homogeneous polynomial of degree $L$
in the $L$ variables $c_0,\,c_1,\,\dots,\,c_{L-1}$.
In order to show that this polynomial does not vanish identically, it suffices to show that
there is at least one monomial in $\det(M)$ with a nonzero coefficient.

Such a monomial, $p_M$, is defined recursively as follows.
If $M$ is $1\times 1$, then $\det(M)$ is a multiple of a single variable $c_j$ and we define $p_M=c_j$.
If $M$ is $d\times d$, let $c_j$ be the variable of lowest index appearing in $M$.
Choose any entry of $M$ in which $c_j$ appears, eliminate from $M$ the row and column containing that entry,
and call the remaining $(d-1)\times(d-1)$ matrix $M'$.  Define $p_M = c_j\,p_{M'}$.

The remainder of the proof  consists of showing that the coefficient
of $p_M$ is nonzero.  In fact, it is a product of minors of $W_L$ which, since $L$ is prime, never vanish due to a
classical result known as Chebotarev's Theorem.

\begin{IEEEproof}(Theorem~\ref{thm:smallsparksampta13})
Let $k\le L$ and choose $k$ columns of $G(c)$.  Applying the algorithm described above to the non-square $L\times k$ matrix $M_0$
formed by those columns, we can identify a monomial $p_{M_0}$.  The key observation is that at each step in the algorithm,
a variable $c_j$ appears for which $0\le j<k$.  Once $k$ rows of $M_0$ have been eliminated, define $M$ to be the $k\times k$
submatrix of $G(c)$ consisting of those rows and the columns of $G(c)$ chosen originally.

Since the polynomial $\det(M)$ is not identically zero, and since at least one nonvanishing monomial of $\det(M)$ has only variables $c_j$
for $0\le j<k$ appearing, there is a $c\in\C^L$, with $\supp(c)\subseteq\set{0,\,\dots,\,k-1}$ such that the columns of $G(c)$
are linearly independent.

Since the exceptional set of such $c$ is the zero set of a polynomial in $k$ variables, its complement is dense and open in $\C^k\times\{0\}$.
Hence the (finite) intersection of these sets over all choices of $k$ columns of $G(c)$ is also dense and open in $\C^k\times\{0\}$.
\end{IEEEproof}

\subsection{Proof of Theorem~\ref{thm:characterization}}

\begin{IEEEproof} Note first that by \eqref{eqn:boldZvector} and \eqref{eqn:boldetavector},
\begin{align*}
 &\sum_{p=0}^L \iint_{[0,T]{\times}[0,\Omega] } |{\bf Z}_{Hg}(t,\nu)_p|^2dt\,d\nu \\ &= \iint_{[0,T]{\times}[0,L\Omega] } |Z_{1/\Omega}Hg(t,\nu)|^2dt\,d\nu= \|Hg\|_{L^2}^2
\end{align*}
and
\begin{align*}
 &\sum_{q,m=0}^L \iint_{[0,T]{\times}[0,\Omega] } |\boldsymbol\eta_H(t,\nu)_{q,m}|^2dt\,d\nu \\&=  \Omega\|\eta^{QP}_H\|_{L^2([0,LT]{\times}[0,L\Omega])}.
\end{align*}

\smallskip\noindent (i)$\Longrightarrow$(iii). If \eqref{eqn:characterization1} fails, then there exist integers $q_0$ and $m_0$ with $S'=S\cap S{+}(m_0LT,n_0L\Omega)$ is a set of positive measure.  This implies that there exists an operator $H\in OPW^2(S)$ with  spreading function $\eta\in L^2(\R)\setminus\{0\}$ and $\eta^{QP}=0$.  Indeed, as $S'\subseteq S{+}(m_0LT,n_0L\Omega)$, we have $S',\,S''=S'{-}(m_0LT,n_0L\Omega)\subseteq S$ and $\eta(t,\nu)=\chi_{S'}(t,\nu)-\chi_{S''}(t,\nu)e^{2\pi i\nu m_0 \nu}\neq 0$ but $\eta^{QP}=0$.
Then ${\bf Z}_{Hg}=0$ which is equivalent to $Hg=0$, showing that (i) fails.

Assume now that \eqref{eqn:characterization1} holds, and, without loss of generality,  $\supp\eta\subseteq {[0,LT]\times [0,L\Omega]}$, so that $\eta^{QP}|_{[0,LT]{\times}[0,L\Omega]}=\eta$.

If \eqref{eqn:characterization2} fails,Then there exists a set of positive measure $A$ with $\sum_{k,\ell} \chi_{S+(kT,\ell\Omega)}(t,\nu)\geq  L+1,\quad (t,\nu)\in A.$

Hence, there exists $\widetilde A\subseteq A$ of positive measure and a fixed collection of $L+1$ rectangles $R_{k,\ell}$ indexed by $\Lambda$ out of the $L^2$ rectangles of size  $[0,T]{\times}[0,\Omega]$ tiling $[0,LT]{\times}[0,L\Omega]$ with $\sum_{(k,\ell)\in \Lambda} \chi_{S+(kT,\ell\Omega)}(t,\nu) \geq  L+1,\quad (t,\nu)\in \widetilde A.$
Since $G(c)|_\Lambda$ has $L+1$ linearly dependent columns, we can choose a nontrivial vector ${\bf x}$ supported on $\Lambda$
with  with ${\bf 0}= G(c){\bf x}$, and, this allows us similarly to above to define a function $\eta_H\neq 0$ supported on $\widetilde A\subseteq S$ with $G(c)\boldsymbol\eta_H(t,\nu)=0$.  As before, we conclude that $Hg=0$ while $H\neq 0$.

\smallskip\noindent (iii)$\Longrightarrow$(ii). Following the arguments above,
$$
	A\|H \|_{HS} \leq \|Hg \|_{L^2}\leq B\|H \|_{HS}
$$
with $A$ is the minimum over all singular values of $L\times L$ sub-matrices of $G(c)$ and $B$ is the maximum over all singular values of $L\times L$ sub-matrices of $G(c)$.

\smallskip\noindent (ii)$\Longrightarrow$(i).  Obvious.
%
%
 \end{IEEEproof}

\subsection{Proof of Theorem~\ref{thm:rectification2}}


\begin{IEEEproof}
Let $L\in\N$ be as described.  We will show that $S$ meets at most $\sigma L$ rectangles $R_{q,m}$, with $T=\sqrt{L}$.
To this end, note that a Jordan curve $C_i$ with length $u_i\in ((k_i-1)/\sqrt{L},k_i/\sqrt{L})$, $k_i\in \N$,
touches at most $4k_i$ rectangles $R_{q,m}$, in fact, this bound is rather pessimistic and only sharp for $k_i=1$. Note that
\begin{align*}
\sqrt{L}U & \geq\sqrt{L} \sum_{i=1}^{N} u_i \\
& \geq\sqrt{L} \sum_{i=1}^{N} (k_i-1)/\sqrt{L}=\big(\sum_{i=1}^{N} k_i\big)-N,
\end{align*}
and, hence, the number of rectangles $B(\partial S)$ needed to cover the boundary $\partial S$ of $S$ satisfies
$$
B(\partial S)\leq  \sum_{i=1}^{N} B(C_i)\leq  \sum_{i=1}^{N} 4 k_i  \leq 4 ( \sqrt{L}U +N).
$$
We conclude that the "fat" boundary, that is, the $1/\sqrt{L}\times 1/\sqrt{L}$ rectification of the boundary has area bounded above by
$$4 ( \sqrt{L}U +N)/(\sqrt{L})^2 = 4 ( U/\sqrt{L} +N/L)\leq \epsilon.$$
It follows immediately, that at most $\sigma L$ sets $R_{q,m}$ are needed to cover $S$.
\end{IEEEproof}

\subsection{Proof of Lemma~\ref{lem:rectification3}}

\begin{IEEEproof}
Note first that with $S^\circ$ given by \eqref{eqn:periodizationofS},
$$\sum_{k,\ell} \chi_{S+(kT,\ell/(TL))} = \sum_{0\le q,m<L} \chi_{S^\circ+(qT,m/(TL))}$$
so that (\ref{eqn:characterization2}) is equivalent to
$$\sup_{(t,\nu)\in[0,T]\times [0,1/(TL)]} \sum_{0\le q,m<L} \chi_{S^\circ+(qT,m/(TL))}(t,\nu) \le L.$$

Assume that \eqref{eqn:characterization2} holds.  Then for each $(t,\nu)\in[0,T]\times [0,1/(TL)]$
there is a unique $0\le n\le L$ and $\abs{\Gamma}=n$ such that
\begin{eqnarray}\label{eqn:sumcondition}
&   & \hskip-.5in \sum_{0\le q,m<L} \chi_{S^\circ+(qT,m/(TL))}(t,\nu) \nonumber \\
& = & \sum_{(q,m)\in\Gamma} \chi_{S^\circ+(qT,m/(TL))}(t,\nu) = n.
\end{eqnarray}
For each such $n$ and $\Gamma$, define the set
$$A_{n,\Gamma} = \set{(t,\nu)\in[0,T]\times [0,1/(TL)]\colon (\ref{eqn:sumcondition})\ \mbox{\rm holds}}.$$
This collection of sets forms the desired partition of $[0,T]\times [0,1/(TL)]$.  It is clear that the sets $S_j$
defined in \eqref{eqn:disjointunion} satisfy the required conditions.

For the other implication, if $A_j+(kT,\ell/(LT))$, $0\le k,\ell<L$, meets $S^\circ$ at most $L$ times, then
\begin{eqnarray*}
&   &  \sup_{(t,\nu)\in[0,T]\times [0,1/(TL)]} \sum_{0\le q,m<L} \chi_{S^\circ+(qT,m/(TL))}(t,\nu) \\
& = & \sup_{(t,\nu)\in[0,T]\times [0,1/(TL)]} \\
&   & \hskip.25in \sup_{1\le j\le N}\sum_{0\le q,m<L} \chi_{A_j+(qT,m/(TL))}(t,\nu) \le L.
\end{eqnarray*}
\end{IEEEproof}

\subsection{Proof of Theorem~\ref{thm:reconstructionquasiperiodic}}

\begin{IEEEproof}
Suppose first that $(t_0,\nu_0)=(0,0)$, and that $c$ is chosen so that $G(c)$ has full spark.
By the support assumption on $S$, (\ref{eqn:matrixequationquasiperiodic})
implies that for $0\le p\le L-1$, \eqref{eqn:basiclinearsystem} takes the form
$${\bf Z}_{Hg}(t,\nu)_p = \sum_{q,m=0}^{L-1} G(c)_{p,(q,m)}\,\boldsymbol\eta_H(t,\nu)_{(q,m)}.$$
Let $[b_{(q,m),p}]$ be a left-inverse of the $L\times\abs{\Gamma}$ matrix $[G(c)_{p,(q,m)}]_{0\le p<L,(q,m)\in\Gamma}$.  That is, for every $(q,m)$, $(q',m')\in\Gamma$,
\begin{equation}\label{eqn:leftinverse}
\sum_{p=0}^{L-1} b_{(q,m),p}\,G(c)_{p,(q',m')} = \frac{1}{\Omega}\,e^{2\pi iqm/L}\delta_{q-q'}\,\delta_{m-m'}.
\end{equation}

Again by support considerations on $S$, $\eta_H\in OPW^2(S)$ satisfies
\begin{align*}
&  \eta^{QP}_H(t,\nu)\CHI_{[0,1/\Omega]}(t)\CHI_{[0,1/T]}(\nu) \\
& \hskip.1in  =  \sum_{(q,m)\in\Gamma} \eta^{QP}_H(t,\nu)\CHI_{[0,T]}(t-qT)\CHI_{[0,\Omega]}(\nu-m\Omega),
\end{align*}
and for each $(q,m)\in\Gamma$,
\begin{align*}
& \eta^{QP}_H(t+qT,\nu+m\Omega)\CHI_{[0,T]}(t)\CHI_{[0,\Omega]}(\nu) \\
& \hskip.1in  =  \sum_{p=0}^{L-1} b_{(q,m),p}\,\CHI_{[0,T]}(t)\CHI_{[0,\Omega]}(\nu)\,e^{2\pi i\nu(q-p)T} \\
& \hskip1in (Z_{1/\Omega}\circ H)g(t+pT,\nu).
\end{align*}
Therefore, by the quasiperiodicity of the Zak transform,
\begin{align*}
&     \eta^{QP}_H(t,\nu)\CHI_{[0,1/\Omega]}(t)\CHI_{[0,1/T]}(\nu) \\
&  =     \sum_{(q,m)\in\Gamma} \sum_{p=0}^{L-1} b_{(q,m),p}\,\CHI_{[0,T]}(t-qT)\CHI_{[0,\Omega]}(\nu-m\Omega) \\
&       \hskip.1in e^{2\pi i(\nu-m\Omega)(q-p)T}\,(Z_{1/\Omega}\circ H)g(t-(q-p)T,\nu).
\end{align*}
Applying (\ref{eqn:reconstructionfromquasiperiodization}),
\begin{align*}
& \eta_H(t,\nu) =    \eta^{QP}_H(t,\nu)\,\CHI_S(t,\nu) \\
&  =    \sum_{k,\ell}  e^{2\pi ik\nu/\Omega}\,\eta^{QP}_H(t-k/\Omega,\nu-\ell/T) \\
& \hskip.1in \big[\CHI_{[0,1/\Omega]}(t-k/\Omega)\CHI_{[0,1/T]}(\nu-\ell/T)\,\CHI_S(t,\nu)\big] \\
&  =    \sum_{k,\ell} e^{2\pi ik\nu/\Omega} \sum_{(q,m)\in\Gamma,p=0}^{L-1} \!\!\!b_{(q,m),p}\,e^{2\pi i(\nu-\ell/T-m\Omega)(q-p)T} \\
& \hskip.1in (Z_{1/\Omega}\circ H)g(t-k/\Omega-(q-p)T,\nu-\ell/T) \\
& \hskip.1in \big[\CHI_{[0,T]}(t-k/\Omega-qT)\CHI_{[0,\Omega]}(\nu-\ell/T-m\Omega)\CHI_S(t,\nu)\big] \\
&  =    \sum_{k,\ell} e^{2\pi ik\nu/\Omega}\,\sum_{(q,m)\in\Gamma,p=0}^{L-1} \!\!\!b_{(q,m),p}\,e^{2\pi i(\nu-m\Omega)(q-p)T} \\
& \hskip.1in e^{-2\pi i\nu k\Omega}\,(Z_{1/\Omega}\circ H)g(t-(q-p)T,\nu) \\
& \hskip.1in \big[\CHI_{[0,T]}(t-k/\Omega-qT)\CHI_{[0,\Omega]}(\nu-\ell/T-m\Omega)\CHI_S(t,\nu)\big] \\
&  =    \sum_{(q,m)\in\Gamma,p=0}^{L-1} \!\!\!b_{(q,m),p}\,e^{2\pi i(\nu-m\Omega)(q-p)T} \\
& \hskip.1in (Z_{1/\Omega}\circ H)g(t-(q-p)T,\nu) \\
& \hskip.25in \biggl[\sum_{k,\ell} \CHI_{[0,T]}(t-k/\Omega-qT) \\
& \hskip.5in \CHI_{[0,\Omega]}(\nu-\ell/T-m\Omega)\,\CHI_S(t,\nu)\biggr].
\end{align*}
Defining
$$S_{(q,m)} = S\cap \biggl(\bigcup_{k,\ell} R_{q,m} + (k/\Omega,\ell/T)\biggr),$$
it follows that $S=\bigcup_{(q,m)\in\Gamma} S_{(q,m)}$, that the union is disjoint, and that
\begin{align*}
\CHI_{S_{(q,m)}}(t,\nu) &= \sum_{k,\ell} \CHI_{[0,T]}(t-k/\Omega-qT) \\
& \hskip.5in \CHI_{[0,\Omega]}(\nu-\ell/T-m\Omega)\,\CHI_S(t,\nu).
\end{align*}
Therefore,
\begin{align*}
& \eta_H(t,\nu) \\
&  =  \sum_{(q,m)\in\Gamma,p=0}^{L-1} b_{(q,m),p}\,e^{-2\pi i(\nu-m\Omega)(q-p)T} \\
& \hskip.1in \sum_{n\in\Z} Hg(t-n/\Omega-(q-p)T)\,e^{2\pi i\nu n/\Omega}\,\CHI_{S_{(q,m)}}(t,\nu) \\
&  =    \sum_{(q,m)\in\Gamma}\bigg[\sum_{p=0}^{L-1} \sum_{n\in\Z} b_{(q,m),p}\,e^{-2\pi i(\nu-m\Omega)(q-p)T} \\
& \hskip.1in  Hg(t-nLT-(q-p)T)\,e^{2\pi i\nu nLT}\,\CHI_{S_{(q,m)}}(t,\nu)\bigg].
\end{align*}
Extending $b_{(q,m),p}$ to have period $L$ in $p$, it follows that
\begin{align*}
& \eta_H(t,\nu) \\
&  =    \sum_{(q,m)\in\Gamma}\bigg[\sum_{p=0}^{L-1} \sum_{n\in\Z} b_{(q,m),p-nL}\,e^{-2\pi i(\nu-m\Omega)(q-(p-nL))T} \\
& \hskip.25in  Hg(t-(q-(p-nL))T)\,\CHI_{S_{(q,m)}}(t,\nu)\bigg] \\
&  =    \sum_{(q,m)\in\Gamma} \sum_k b_{(q,m),k}\,e^{-2\pi i(\nu-m\Omega)(q-k)T} \\
& \hskip.25in  Hg(t-(q-k)T)\,\CHI_{S_{(q,m)}}(t,\nu).
\end{align*}
Finally, writing
$$h(x,t)  =    \int \eta(t,\nu)\, e^{2\pi i (x-t) \nu }\, d\nu$$
yields (\ref{eqn:reconstructionformulaperiodic}) with $(t_0,\nu_0)=(0,0)$.

To complete the proof, note that for almost every $t$, the set, $\set{\nu\colon (t,\nu)\in S_{(q,m)}}$ is contained in a
fundamental domain of the lattice $T\Z$ of $\R$.  This implies that the measure of each such section is no more than $1/T$,
and in particular that for almost every $t$, $\CHI_{S_{(q,m)}}(t,\cdot)\in L^2(\R)$.  Therefore, by Plancherel's Formula,
\begin{align*}
& \int\!\!\!\int\abs{\Phi_{(q,m)}(t,s)}^2\,dt\,ds \\
& \hskip.1in = \int\!\!\!\int\biggl|\int e^{2\pi i\nu s}\,\CHI_{S_{(q,m)}}(t,\nu)\,d\nu\biggr|^2\,ds\,dt \\
& \hskip.1in = \int\!\!\!\int\abs{\CHI_{S_{(q,m)}}(t,\nu)}^2\,d\nu\,dt = \abs{S_{(q,m)}}^2 <\infty
\end{align*}
and for almost every $(t,s)$,
$$\abs{\Phi_{(q,m)}(t,s)} \le  \int \CHI_{S_{(q,m)}}(t,\nu)\,d\nu \le 1/T.$$
Hence $\Phi_{(q,m)}\in L^2\cap L^\infty(\tfR)$.  Convergence of the reconstruction sum in $L^2(\R^2)$ follows from the observation that
$Hg\in L^2(\R)$ (see Lemma~\ref{lem:deltatrain}) and basic properties of the Zak Transform (see e.g., \cite{Gro01}, Section~8.2).

If $(t_0,\nu_0)\neq (0,0)$, we formally compute
\begin{align*}
  H &=  \iint_S \eta_H(t,\nu)\, M_\nu T_t \ dt\,d\nu \\
    &=   \iint_{ S-(t_0,\nu_0)} \eta_H(t+t_0,\nu+\nu_0)\  T_{t+t_0}M_{\nu+\nu_0}  \ dt\,d\nu \\
    &= T_{t_0} M_{\nu_0} \widetilde H,
\end{align*}
where $\eta_{\widetilde H}(t,\nu)=e^{-2\pi i t \nu_0}\,\eta_H(t+t_0,\nu+\nu_0)$.
Taking inverse Fourier transforms $\nu\to x$ on both sides, we obtain
$h_{\widetilde H}(t,x)=e^{-2\pi i t \nu_0}\,h_H(t+t_0,x)\, e^{-2\pi i \nu_0x}$ which is
\begin{equation}\label{eqn:shiftedpointspread}
h_H(t,x)=e^{2\pi i (x+t-t_0) \nu_0}\,h_{\widetilde H}(t-t_0,x).
\end{equation}
With $\widetilde S= S-(t_0,\nu_0)$, we can apply (\ref{eqn:reconstructionformulaperiodic}) with $(t_0,\nu_0)=(0,0)$ to reconstruct
$h_{\widetilde H}$ from $\widetilde H g$ with the same $g=\sum c_n \delta_{nT}$, that is,
\begin{align}
& \hskip.1in  h_{\widetilde H}(x,t) \notag \\
&  =   \sum_k \sum_{(q,m)\in\Gamma} b_{(q,m),k}\,\widetilde Hg(t - (q-k)T)\,e^{-2\pi i m(q-k)/L} \notag \\
& \hskip1in \widetilde \Phi_{(q,m)}(t,(x-t)+(q-k)T). \label{eqn:reconstructionformulaperiodicshifted}
\end{align}
where
$$\widetilde \Phi_{(q,m)}(t,s) = \int e^{2\pi i\nu s}\,\CHI_{\widetilde S_{(q,m)}}(t,\nu)\,d\nu$$
and
$$\widetilde S_{(q,m)} = \widetilde S\cap \bigcup_{k,\ell\in\Z} (\widetilde R_{q,m}+(k/\Omega,\ell/T)).$$
Observing that $S_{(q,m)} = \widetilde S_{(q,m)}+(t_0,\nu_0)$, we obtain
$$\Phi_{(q,m)}(t,s) =e^{2\pi i s \nu_0}\, \widetilde \Phi_{(q,m)}(t-t_0,s).$$
combining \eqref{eqn:reconstructionformulaperiodicshifted} with \eqref{eqn:shiftedpointspread} yields
\begin{align*}
& h(x,t) \\
& \hskip.1in \sum_k \sum_{j=0}^{L-1} b_{(q,m),k}\, (M_{-\nu_0} T_{-t_0}H)g(t-t_0 - (q-k)T) \\
& \hskip.25in \,e^{-2\pi i m(q-k)/L}\,e^{-2\pi i (x-t-t_0+(q-k)T)\nu_0} \\
& \hskip.5in \Phi_{(q,m)}(t,(x-t-t_0)+(q-k)T) \\
&  =   e^{2\pi i (x+t-t_0)\nu_0} \\
& \hskip.1in \sum_k \sum_{(q,m)\in\Gamma} b_{(q,m),k}\, e^{-2\pi i ( t-t_0 - (q-k)T) \nu_0} Hg(t - (q-k)T) \\
& \hskip.25in e^{-2\pi i m(q-k)/L}\,e^{-2\pi i (x-t-t_0+(q-k)T)\nu_0} \\
& \hskip.5in \Phi_{(q,m)}(t,(x-t-t_0)+(q-k)T) \\
&  =   e^{2\pi i (t+t_0)\nu_0} \\
& \hskip.1in \sum_k \sum_{(q,m)\in\Gamma} b_{(q,m),k}\,Hg(t - (q-k)T)\,e^{-2\pi i m(q-k)/L} \\
 &\hskip.5in  \Phi_{(q,m)}(t,(x-(t+t_0)+(q-k)T)\,.
\end{align*}
\end{IEEEproof}

\subsection{Outline of Proof of Theorem~\ref{thm:reconstruction}.}

The proof follows that of Theorem~\ref{thm:reconstructionquasiperiodic} once we establish that we can
replace the sharp cut-offs, $\CHI_{[0,T]}$ and $\CHI_{[0,\Omega]}$ by smooth ones.
Since $S$ is compact and $|S|<1$, for $\delta>0$ sufficiently small, the set
$S_\delta = S + [-\delta,\delta]^2$ also satisfies $|S_\delta|<1$.  Since Theorem~\ref{thm:reconstructionquasiperiodic}
allows us to shift the region, and since $|S|<1$, we can assume without loss of generality that
there exist $T>0$ and $L\in\N$ such that $S\subseteq (0,TL)\times (0,1/T)$ and that $S_\delta$
has a $(T,L)$-rectification.  Since $S\subseteq \cup_{(q,m)\in\Gamma} R_{q,m}=R$,
it is sufficient to prove the theorem with $OPW^2(S)$ replaced by $OPW^2(R)$.

By Lemma~\ref{lem:matrixequationquasiperiodic}, given $H\in OPW^2(R)$ with spreading function $\eta_H(t,\nu)$,
and given any weighted delta train of the form $g=\sum_n c_n\,\delta_{nT}$
where $c=(c_n)$ is a period-$L$ sequence,
(\ref{eqn:matrixequationquasiperiodic})
holds with $\eta^{QP}_H$ replaced by $\eta_H$ for all $(t,\nu)$ in an $\epsilon$-neighborhood of $[0,T]{\times}[0,\Omega]$,
$R_{0,0}^\epsilon=([-\epsilon/2,T+\epsilon/2]{\times}[-\epsilon/2,\Omega+\epsilon/2]$.

Let $r,\varphi\in\mathcal{S}(\R)$ satisfy
\begin{align}\label{eq:r_phi_1}
\supp r & \subseteq  [-\epsilon/2,T+\epsilon/2], \\
\supp \widehat{\varphi} & \subseteq  [-\epsilon/2,\Omega+\epsilon/2], \nonumber
\end{align}
so that $\supp r(t)\widehat{\varphi}(\nu) \subseteq R_{0,0}^\epsilon$, and
\begin{equation}\label{eq:r_phi_2}
\sum_{k\in\Z} r(t + kT) = 1 = \sum_{n\in\Z} \widehat{\varphi}(\nu + n\Omega),
\end{equation}
for all $(t,\nu)\in\tfR$.  For $\epsilon<\delta$, it is not hard to show that
if $R_{q,m}\not\subseteq R$ then
\begin{equation}\label{eq:r_phi_3}
\eta_H(t,\nu) r(t-qT) \widehat{\varphi}(\nu-m\Omega) = 0.
\end{equation}
Therefore,
$$\eta_H(t,\nu) = \sum_{(q,m)\in\Gamma} \eta^{QP}_H(t,\nu)\,r(t-qT)\,\widehat{\varphi}(\nu-m\Omega).$$
Following the proof of Theorem~\ref{thm:reconstructionquasiperiodic}, with $r(t)$ replacing $\CHI_{[0,T]}(t)$ and
$\widehat{\varphi}(\nu)$ replacing $\CHI_{[0,\Omega]}(\nu)$,
\begin{align*}
\eta_H(t,\nu) = & \sum_{(q,m)\in\Gamma} \sum_k b_{(q,m),k}\,e^{-2\pi i(\nu-m\Omega)(q-k)T} \\
& \hskip.5in  Hg(t-(q-k)T)\,R_{(q,m)}(t,\nu)
\end{align*}
where
\begin{align*}
& R_{(q,m)}(t,\nu) \\
& \hskip.1in =    \sum_{k,\ell} r(t-k/\Omega-qT)\widehat{\varphi}(\nu-\ell/T-m\Omega)\,\CHI_R(t,\nu) \\
& \hskip.1in =    r(t-qT)\,\widehat{\varphi}(\nu-m\Omega).
\end{align*}
Finally,
\begin{align*}
& h(x,t) \\
& = \sum_{(q,m)\in\Gamma} \sum_k b_{(q,m),k}\,e^{2\pi i m(q-k)/L}\,Hg(t-(q-k)T)\\
& \hskip.5in \Phi_{(q,m)}(t,(x-t) + (q-k)T)
\end{align*}
where here
\begin{align*}
\Phi_{(q,m)}(t,s)
&  =    \int e^{2\pi i\nu s}\,R_{(q,m)}(t,\nu)\,d\nu \\
&  =    r(t-qT)\,e^{2\pi i s m\Omega}\varphi(s).
\end{align*}
Plugging this into (\ref{eqn:reconstructionformulaperiodic}) gives the result.

\subsection{Lattice tilings and proof of Theorem~\ref{thm:reconstructionquasiperiodicsymplectic}}
\label{sec:reconstructionquasiperiodicsymplectic}

In this section we will prove Theorem~\ref{thm:reconstructionquasiperiodicsymplectic},
but also derive results where the tiling of $S$ is defined by arbitrary full rank lattices in $\R^2$.
The  reconstruction formulas  use results from representation theory; these carry over to the
higher dimensional setting if the lattice is symplectic.

\begin{IEEEproof}
As before, we assume that $S\subseteq\tfR$ satisfies $\abs{S}<1$. Suppose that for some
$A=
\big(\begin{smallmatrix}
 a_{11} & a_{12} \\ a_{21} & a_{22}
\end{smallmatrix}\big)$ with $\det A=1/L$, $S$ is contained in a fundamental
domain of the lattice $LA \Z^2$.
The lattice $LA \Z^2$  is the so-called adjoint lattice $A^\circ$ of $A$. Indeed,
$A^\circ= (1/\sqrt{L}) \ (\sqrt{L}\,A)^\circ = \sqrt{L} \sqrt{L}A=LA$ (see \cite{Gro01} for details).
We shall assume without loss of generality that $a_{11}\neq0$.
Otherwise, we could replace the first column with the second and the second with the negative of the first,
leading to a different parametrization of the same lattice.
Further assume that there exist $t_0,\nu_0$, and $\Gamma\subseteq\Z^2$, $\abs{\Gamma}\le L$
such that with $P_{q,m} = A \big( [0,1]^2{+} (t_0,\nu_0) +(q,m)^T\big)$, $q,\,m\in\Z$,
\begin{equation}\label{eqn:rectificationofperiodizedsupport3}
LA [0,1]^2\cap\bigg(\bigcup_{k,\ell\in\Z} S+ LA  (k,\ell)^T\bigg) \subseteq \bigcup_{(q,m)\in\Gamma} P_{q,m}.
\end{equation}
As before, we will set
$$\Phi_{(q,m)}(t,s) = \int e^{2\pi i\nu s}\,\CHI_{S_{(q,m)}}(t,\nu)\,d\nu$$
where
$$S_{(q,m)} = S\cap \bigcup_{k,\ell\in\Z} (P_{q,m}+LA  (k,\ell)^T).$$

We will derive reconstruction formulas and show  that if $a_{12}/a_{11}$ is rational,
then $OPW^2(S)$ can be identified with a weighted delta train and if
$a_{21} a_{11}$ is rational as well, then the coefficient sequence
$(\widetilde {c}_n)$ of that delta train is periodic and we are in the framework of regular operator sampling.

We shall assign to each operator $H\in OPW^2(S)$ an operator in $\widetilde H\in OPW^2(L^{-1/2}A^{-1} S )$
and then apply the reconstruction formula in Theorem~\ref{thm:reconstructionquasiperiodic} to reconstruct $\widetilde h=h_{\widetilde H}$
of $\widetilde H \in OPW^2(L^{-1/2}A^{-1} S)$. From this, we will construct $h=h_H$ and therefore $H$.

The result is based on the existence of the  operators $\mu(\sqrt{L}A)$ that appear in the following computation.
The existence follows from the representation theory of the Weyl-Heisenberg group and is discussed in this setting in
\cite{KP06,Pfa10}.  Let  $\rho(t,\nu)=e^{\pi i t \nu}T_t M_\nu$, $\eta^\#(t,\nu)=e^{-\pi i t \nu} \eta(t,\nu)$, and  $B=\sqrt{L}A$. Then
\begin{align*}
H 	&=  \iint \eta(t,\nu)\,  T_t M_\nu\ dt\,d\nu \\
     	&=  \iint_S \eta(t,\nu)e^{-\pi i t \nu}\,  e^{\pi i t \nu}T_t M_\nu\ dt\,d\nu \\
    	&=  \iint_S \eta^\#(t,\nu)\  \rho(t,\nu) \,dt\,d\nu \notag\\
    	&=   \iint_{B^{-1}(S)} \eta^\#(B (t,\nu))\  \rho(B (t,\nu)) \ dt\,d\nu \\
    	&=  \iint \eta^\#( B (t,\nu))\ {\mu(B)} \rho(t,\nu)\,\mu(B)^{\ast}  \ dt\,d\nu \notag\\
    	&=  {\mu(B)}\iint\eta^\#( B (t,\nu))\  \rho(t,\nu) \ dt\,d\nu\ \mu(B)^{\ast}\\
    	&=  {\mu(B)}\ \widetilde H \ \mu(B)^{\ast}\notag\,,
\end{align*}
with $\widetilde \eta^\#(t,\nu)=\eta^\# (B(t,\nu))$. Setting $Q_1(t,\nu)=t$ and $Q_2(t,\nu)=\nu$ we have
$$\widetilde \eta(t,\nu)=e^{\pi i (t \nu - Q_1B(t,\nu)\cdot Q_2B(t,\nu) )} \eta (B(t,\nu)).$$

Moreover, observe that $\widetilde S= B^{-1}S$ satisfies the hypothesis of Theorem~\ref{thm:reconstructionquasiperiodic}
with $T=\Omega=1/\sqrt{L}$.  We have therefore with an $L$ periodic sequence $(\widetilde c_n)$,
$\widetilde g=\sum \widetilde c_n \delta_{n\sqrt L}$, and $B^{-1}=\big(\begin{smallmatrix}
 b_{22} & -b_{12} \\ -b_{21} & b_{11}
\end{smallmatrix}\big)$ the reconstruction formulas
\begin{align*}
 \widetilde h(x,t) &= e^{2\pi i (t+t_0)\nu_0}\sum_k \sum_{(q,m)\in\Gamma} \big[b_{(q,m),k} \notag \\
& \hskip-.25in\widetilde H \widetilde g (t - (q-k)/\sqrt{L}) e^{-2\pi i m(q-k)/L}  \notag \\
& \widetilde \Phi_{(q,m)}(t{+}t_0,x-(t{+}t_0)+(q-k)/\sqrt{L})\big], \notag
\end{align*}
\begin{align*} \widetilde \eta (t,\nu) &= e^{2\pi i (t+t_0)\nu_0}\sum_k \sum_{(q,m)\in\Gamma} \big[b_{(q,m),k}  \notag \\
& \hskip-.25in \widetilde{H} \widetilde g (t - (q-k)/\sqrt{L}) e^{-2\pi i m(q-k)/L}  \notag \\
& \chi_{B^{-1}S_{(q,m)}}(t{+}t_0,\nu) e^{2\pi i (t{+}t_0- (q-k)/\sqrt{L})\nu}\big] \notag \\
&= e^{\pi i (t \nu - Q_1B(t,\nu)\cdot Q_2B(t,\nu) )}
		\eta (B(t,\nu)) \notag
\end{align*}
\begin{align}
 \eta (t,\nu) &= e^{2\pi i (Q_1B^{-1}(t,\nu)+t_0)\nu_0}  \notag \\
& \hskip-.25in	e^{-\pi i ( Q_1B^{-1}(t,\nu)\cdot Q_2B^{-1}(t,\nu) -t\nu)}
 		\sum_k \sum_{(q,m)\in\Gamma} \big[b_{(q,m),k}  \notag \\
& \hskip-.25in	\widetilde{H} \widetilde g (Q_1B^{-1}(t,\nu) - (q-k)/\sqrt{L})  \notag \\
& \hskip-.25in e^{-2\pi i m(q-k)/L} \chi_{S_j}\big((t,\nu){+}B(t_0,0)\big)  \notag \\
& \hskip-.25in e^{2\pi i (Q_1B^{-1}(t,\nu)+t_0- (q-k)/\sqrt{L})Q_2B^{-1}(t,\nu)}\big] \notag \\
&  =   	e^{2\pi i ((b_{22}t-b_{12}\nu)\nu_0+t_0(b_{11}\nu-b_{21}t) +t_0\nu_0)}  \notag \\
& \hskip-.25in e^{\pi i ( (b_{22}t-b_{12}\nu)\cdot (b_{11}\nu-b_{21}t) -t\nu)}\sum_k \sum_{(q,m)\in\Gamma} \big[b_{(q,m),k}  \notag \\
& \hskip-.25in \widetilde{H} \widetilde g ((b_{22}t-b_{12}\nu) - (q-k)/\sqrt{L})  \notag \\
& \hskip-.25in	e^{-2\pi i m(q-k)/L}\chi_{S_{(q,m)}}\big((t,\nu){+}(b_{11}t_0, b_{21}t_0)\big)  \notag \\
& \hskip-.25in	e^{-2\pi i  (q-k)/\sqrt{L})(b_{11}\nu-b_{21}t)}\big]\notag \\
&  =  e^{2\pi i \big([(a_{22}t-a_{12}\nu)\nu_0+t_0(a_{11}\nu-a_{21}t)]\sqrt{L} +t_0\nu_0\big)}  \notag \\
& \hskip-.25in e^{\pi i (L (a_{22}t-a_{12}\nu)\cdot (a_{11}\nu-a_{21}t) -t\nu)}\sum_k \sum_{(q,m)\in\Gamma} \big[b_{(q,m),k}  \notag \\
& \hskip-.25in \widetilde{H} \widetilde g ((a_{22}t-a_{12}\nu)\sqrt{L} - (q-k)/\sqrt{L})\notag \\
& \hskip-.25in e^{-2\pi i m(q-k)/L}\chi_{S_{(q,m)}}(t + \sqrt{L}a_{11}t_0,\, \nu+ \sqrt{L}a_{21}t_0)  \notag \\
& \hskip-.25in e^{-2\pi i  (q-k)(a_{11}\nu-a_{21}t)}\big]. \label{eqn:reconsymplecticgeneral}
 \end{align}
Taking inverse Fourier transforms  $\nu\to x$ on both sides gives us a formula for $h$,
but as the right hand side contains the product of three functions in $\nu$,
the resulting formula for $h$ does not give much insight in general. If $a_{12}=0$ though,
the above simplifies (using $a_{11} a_{22}=1/L$) to
\begin{align}
 \eta(t,\nu)
    &= \sum_k \sum_{(q,m)\in\Gamma} \big[b_{(q,m),k}\,
		\widetilde{H} \widetilde g (a_{22}\sqrt{L}t - (q-k)/\sqrt{L}) \notag \\
 &	\hskip-.25in	e^{-2\pi i m(q-k)/L}\,\chi_{S_{(q,m)}}(t + \sqrt{L}a_{11}t_0,\, \nu+ \sqrt{L}a_{21}t_0) \notag \\
 &	\hskip-.25in	e^{2\pi i (t_0\nu_0 a_{11}\sqrt{L} -(q-k)a_{11})\nu}\,
		e^{-2\pi i (\sqrt{L}t_0\nu_0 +L/2) a_{22}a_{21}t^2} \notag \\
 &	\hskip-.25in	e^{2\pi i t_0\nu_0} \, e^{2\pi i (q-k)a_{21}t}\big] \label{eqn:reconsymplecticgeneral01}
\end{align}
which leads to
\begin{align}
h(x,t) & =
		e^{-2\pi i (\sqrt{L}t_0\nu_0 +L/2) a_{22}a_{21}t^2}\,
		e^{2\pi i t_0\nu_0}  \notag  \\
&  \sum_k \sum_{(q,m)\in\Gamma} \big[b_{(q,m),k} \widetilde{H} \widetilde g (\sqrt{L}(a_{22}t- (q-k)/L)) \notag \\
&  e^{-2\pi i m(q-k)/L} \notag \\
&  \Phi_{(q,m)}(t + \sqrt{L}a_{11}t_0t, \, x +t_0\nu_0 a_{11}\sqrt{L} -(q-k)a_{11}) \notag \\
&		e^{-2\pi i \sqrt{L} a_{21}t_0(x+t_0\nu_0 a_{11}\sqrt{L} -(q-k)a_{11})}  \,
		e^{2\pi i (q-k)a_{21}t}\big]  \label{eqn:reconsymplecticgeneral02}
\end{align}
and, if $t_0=0$,
\begin{align}
h(x,t) & =
		e^{-\pi i L a_{22}a_{21}t^2} \notag \\
& \hskip-.35in		\sum_k \sum_{(q,m)\in\Gamma} \big[b_{(q,m),k}\,
		\widetilde{H} \widetilde g (\sqrt{L}(a_{22}t- (q-k)/L)) \notag \\
&		e^{-2\pi i m(q-k)/L} \,e^{2\pi i (q-k)a_{21}t} \notag \\
& \hskip.35in \Phi_{(q,m)}(t , \, x  -(q-k)a_{11})\big]  \label{eqn:reconsymplecticgeneral03}
\end{align}

By construction, we have $\widetilde H\widetilde g=\mu(B)^\ast H \mu(B)\widetilde g$ with $\widetilde g=\sum \widetilde c_n \delta_{n/\sqrt{L}}$.
Hence, we can replace $\widetilde H$ in \eqref{eqn:reconsymplecticgeneral} by $\mu(B)^\ast H$ and  $\widetilde g$ by $g$ where $g=\mu(B)\widetilde g$.
In the following, we will give explicit representation of $\mu(B)$ and examine  $g=\mu(B)\widetilde g$.
Note that the given reconstruction formulas hold true for any tempered distribution $g=\mu(B)\widetilde g$,
but we are mainly interested in the case that $\mu(B)\widetilde g$ is discretely supported, or, better,
$g=\mu(B)\widetilde g=\sum \widetilde c_n \delta_{n  T}$ for some $T>0$ and a periodic sequence $c=(c_n)$.
In applications, this would allow us to use any hardware developed to excite an operator described in Theorem~\ref {thm:reconstructionquasiperiodic}.

Recall that  $B=\sqrt{L}A$, so $\det B=1$ and we assume   $b_{11}\neq 0$.
We have
\begin{align}
& \hskip.25in \big(\begin{smallmatrix}
 b_{11} & b_{12} \\ b_{21} & b_{22}
\end{smallmatrix}\big) \notag \\
&=
 \big(\begin{smallmatrix}
 1& 0 \\ b_{21}/b_{11} & 1
\end{smallmatrix}\big)\big(\begin{smallmatrix}
 0 & -1 \\ 1 & 0
\end{smallmatrix}\big)\big(\begin{smallmatrix}
 1& 0 \\ -b_{11}b_{12} & 1
\end{smallmatrix}\big)\big(\begin{smallmatrix}
 0 & 1 \\ -1 & 0
\end{smallmatrix}\big)\big(\begin{smallmatrix}
 b_{11} & 0 \\ 0 & 1/b_{11}
\end{smallmatrix}\big)
\end{align}

Using notation from \cite{Gro01}, we have
\begin{align*}
\mu_1(\alpha)&=\mu
  	\big(\begin{smallmatrix}
 		1& 0 \\ \alpha & 1
	\end{smallmatrix}\big)
	: f \mapsto  e^{\pi i \alpha (\cdot)^2} f\,, \\		
\mathcal F&=\mu
	\big(\begin{smallmatrix}
 		0 & 1 \\ -1 & 0
	\end{smallmatrix}\big)
	: f \mapsto \widehat f\,,\\
\mu_2(\alpha)&=\mu\big(\begin{smallmatrix}
 		\alpha & 0 \\ 0 & 1/\alpha
	\end{smallmatrix}\big)
	: f \mapsto \mathcal  \alpha^{-1/2} f( \, \cdot\,/ \alpha),
\end{align*}
hence,
\begin{align*}
\mu\big(\begin{smallmatrix}
 b_{11} & b_{12} \\ b_{21} & b_{22}
\end{smallmatrix}\big)
&= \mu_1(b_{21}/b_{11})\,  \mathcal{F}^\ast  \mu_1(-b_{11}b_{12}) \,\mathcal F \, \mu_2(b_{11}) \\
& \hskip-.35in = \mu_1(a_{21}/a_{11})\,  \mathcal{F}^\ast  \mu_1(-La_{11}a_{12}) \,\mathcal F \, \mu_2(\sqrt{L}a_{11}).
\end{align*}
This leads to
\begin{align*}
\mu(B)\widetilde g&=
\mu\big(\begin{smallmatrix}
 b_{11} & b_{12} \\ b_{21} & b_{22}
\end{smallmatrix}\big) \sum c_n \delta_{n/\sqrt{L}}
 \\& = \mu_1(a_{21}/a_{11})\,  \mathcal{F}^\ast  \mu_1(-La_{11}a_{12}) \,\mathcal F \circ \\
   & \hskip.5in  \mu_2(\sqrt{L}a_{11})\sum c_n \delta_{n/\sqrt{L}} \\
 \\& = (\sqrt{L}a_{11})^{-1/2}\  \mu_1(a_{21}/a_{11})\,  \mathcal{F}^\ast  \mu_1(-La_{11}a_{12}) \circ \\
   & \hskip.5in  \mathcal F \sum c_n \delta_{na_{11}} \\
 \\& = (\sqrt{L}a_{11})^{-1/2}\  \mu_1(a_{21}/a_{11})\circ \\
   & \hskip.5in  \mathcal{F}^\ast  \mu_1(-La_{11}a_{12}) \sum \widehat{c}_m \delta_{m/(L a_{11})} \\
 \\& = (\sqrt{L}a_{11})^{-1/2}\  \mu_1(a_{21}/a_{11})\circ \\
   & \hskip.5in  \mathcal{F}^\ast
 		\sum \widehat{c}_m \, e^{-2\pi i  m^2 a_{12} /(2 L a_{11})}\delta_{m/(L a_{11})}
 \end{align*}
where we have used the fact that the Fourier transform of a delta train of the form
$\sum_{n\in\Z} c_n\delta_{nT}$, where $c=(c_n)$ has period $L$ is another delta train of
the same form.  Specifically,
\begin{equation}\label{eqn:transformofweighteddeltatrain}
{\cal F}\sum_{n\in\Z} c_n\delta_{nT} = \frac{1}{LT}\sum_{m\in\Z} \widehat{c}_m\,\delta_{m/LT}
\end{equation}
where $\widehat{c}$ denotes the Discrete Fourier Transform of $c$, that is
$$\widehat{c}_m = \sum_{k=0}^{L-1}c_k\,e^{-2\pi ikm/L}.$$
Equation (\ref{eqn:transformofweighteddeltatrain}) is a simple consequence of the fact that
$${\cal F}\sum_{n\in\Z} \delta_{nW} = \frac{1}{W}\,\sum_m \delta_{m/W}.$$
The sequence $e^{-2\pi i  m^2 a_{12} /(2 L a_{11})}$ is periodic in $m$ if $e^{-2\pi i  m a_{12} /(2 L a_{11})}$ is, that is,
if $a_{12} / a_{11}$ is rational.
In the following,  ${\rm LCM}$ refers to  least common multiples of natural numbers,
and for a rational number $a$, ${\rm q}[a]$ denotes the smallest natural number $q$ such that $qa$ is an integer.
With this notation, $(\widehat{c'})_m=\widehat{c}_m \, e^{-2\pi i  m a_{12} /(2 L a_{11})}$ forms a sequence with period
$L'={\rm LCM}\{{\rm q}[a_{12} /(2 L a_{11})],\, L\}$. Once again employing (\ref{eqn:transformofweighteddeltatrain}),
\begin{align}\label{eqn:nonperiodicidentifier}
\notag
& \mu(B) \widetilde g=
 (\sqrt{L}a_{11})^{-1/2}\  \mu_1(a_{21}/a_{11})\,  \mathcal{F}^\ast
 		\sum (\widehat{c'})_m  \delta_{m/(L a_{11})}\\
 \notag
		&=
 (\sqrt{L}a_{11})^{-1/2}\  \mu_1(a_{21}/a_{11})\,
 		\sum c'_n  \delta_{n a_{11} L /L'}\\
		&=
 (\sqrt{L}a_{11})^{-1/2}\
 		\sum c'_n \,  e^{2\pi i n^2 a_{21} a_{11} (L/L')^2/2} \delta_{n a_{11} L /L'}.
 \end{align}
We conclude that $\mu(B) g=\sum \widetilde{c}_n \delta_{nT}$ with $T=a_{11}L/ {\rm q}[a_{12} /(2 L a_{11})]$ if $a_{12} / a_{11}$ is rational.
Moreover, if $a_{21} a_{11}$ is rational as well,  then we are assured that the coefficient sequence  $(\widetilde {c}_n)$
has period
\begin{align*}
L''& ={\rm LCM}\{ {\rm q}[a_{21} a_{11} (L/L')^2/2],\, L'\} \\
   & ={\rm LCM}\{{\rm q}[a_{21} a_{11} (L/{\rm q}[a_{12} /(2 L a_{11})])^2/2], \\
   & \hskip1in {\rm q}[a_{12} /(2 L a_{11})],\, L\},
\end{align*}
that is, we are in the framework of regular operator sampling.

Let us consider the special case that $a_{12}/(2a_{11})$ is an integer
(for example, if $a_{12}=0$ as in Theorem~\ref{thm:reconstructionquasiperiodicsymplectic}),
then ${\rm q}[a_{12} /(2 L a_{11})]\in\{1,L\}$, so  $L'=L$ and
$L''= {\rm LCM}\{{\rm q}[a_{21} a_{11} /2],\,  L\}.$ If in addition $La_{21} a_{11} /2$ is an integer, then
${\rm q}[a_{21} a_{11} /2]\in\{1,L\}$ and $L''=L$.

To complete the proof of Theorem~\ref{thm:reconstructionquasiperiodicsymplectic}, observe
first that  $L=L'$, and indeed $(c_n)=(c'_n)$. Consequently
\begin{align*}
g= \mu(B) \widetilde g&=
 		\sum c_n \,  e^{\pi i n^2 a_{21} a_{11}  } \delta_{n a_{11} }.
 \end{align*}
 Further, observe that
\begin{align*}
& \mu\big(\begin{smallmatrix}
 b_{11} & b_{12} \\ b_{21} & b_{22}
\end{smallmatrix}\big)^\ast \\
 &= \mu_2(\sqrt{L}a_{11})^\ast\mathcal F^\ast   \mu_1(-La_{11}a_{12})^\ast \mathcal{F}\,  \mu_1(a_{21}/a_{11})^\ast
 \\& =  \mu_2(1/(\sqrt{L}a_{11}))\,\mathcal F^\ast   \mu_1(La_{11}a_{12})  \mathcal{F}\,  \mu_1(-a_{21}/a_{11}).
\end{align*}
Hence, if $a_{12}=0$, then
\begin{align*}
& \mu\big(\begin{smallmatrix}
 b_{11} & 0 \\ b_{21} & b_{22}
\end{smallmatrix}\big)^\ast f (x) \\
   &=  \mu_2(1/(\sqrt{L}a_{11}))\,  \mu_1(-a_{21}/a_{11})f(x) \\
   &	= (\sqrt{L}a_{11})^{1/2}\,e^{-\pi i a_{21}/a_{11}  (\sqrt{L}a_{11}x)^2}\, f(\sqrt{L}a_{11}x)
\\& =(\sqrt{L}a_{11})^{1/2}\,e^{-\pi i L a_{21}a_{11}x^2}\, f(\sqrt{L}a_{11}x)
\end{align*}
and
\begin{align*}
& \mu(B)^\ast Hg(\sqrt{L}(a_{22}t- (q-k)/L)) \\
& = (\sqrt{L}a_{11})^{1/2} \,e^{-\pi i L a_{21}a_{11}(\sqrt{L}(a_{22}t- (q-k)/L))^2 } \\
& \hskip.5in H g(\sqrt{L}a_{11} \sqrt{L}(a_{22}t- (q-k)/L)) \\
& = (\sqrt{L}a_{11})^{1/2}\, e^{-\pi i  a_{21}a_{11}(La_{22}t- (q-k))^2 } \\
& \hskip.5in H g(t- a_{11} (q-k))
\end{align*}
We conclude that
\begin{align}
& h(x,t)  =
		(\sqrt{L}a_{11})^{1/2}
		e^{-2\pi i (\sqrt{L}t_0\nu_0 +L/2) a_{22}a_{21}t^2}\,
		e^{2\pi i t_0\nu_0}
\notag \\ &
		\sum_k \sum_{(q,m)\in\Gamma} \big[b_{(q,m),k}\,
		e^{-\pi i  a_{21}a_{11}(La_{22}t- (q-k))^2 } \notag \\
&		\hskip.25in H g(t- a_{11} (q-k))\,
		e^{-2\pi i m(q-k)/L}\,
\notag \\ &   \hskip.25in
		\Phi_{(q,m)}(t + \sqrt{L}a_{11}t_0t, \, x +t_0\nu_0 a_{11}\sqrt{L} -(q-k)a_{11}) \notag \\
&		\hskip.25in e^{-2\pi i \sqrt{L} a_{21}t_0(x+t_0\nu_0 a_{11}\sqrt{L} -(q-k)a_{11})} e^{2\pi i (q-k)a_{21}t} \big]
               \label{eqn:reconsymplecticgeneral04}
\end{align}
and, if $t_0=0$,
\begin{align}
& h(x,t)  =
		(\sqrt{L}a_{11})^{1/2}
		e^{-\pi i L a_{22}a_{21}t^2} \notag \\
&		\sum_k \sum_{(q,m)\in\Gamma} b_{(q,m),k}\,
		e^{-\pi i  a_{21}a_{11}(La_{22}t- (q-k))^2 } \notag \\
& \hskip.25in H g(t- a_{11} (q-k))
\notag \\ & \qquad
		\Phi_{(q,m)}(t , \, x  -(q-k)a_{11}) \,
	  	e^{2\pi i (q-k)a_{21}t} \,.  \label{eqn:reconsymplecticgeneral05}
\end{align}
\end{IEEEproof}

\subsection{Proof of assertion in Example~\ref{exa:non periodic}.}\label{sec:parallelograms}

The goal is to show that $OPW^2(S)$ where $S$ is the region shown in Figure~\ref{fig:nonperiodic} cannot be identified by regular
operator sampling for any $T$ or $L$.  We will show that the $T\Z\times(1/LT)\Z$ periodization of $S$ does not form an exact $L$-cover
for any $T$ or $L$, thus violating \eqref{eqn:characterization2} and Remark~\ref{rem:characterization}(2)

\begin{IEEEproof}
Assume first that $T$ is rational.  We can assume without loss of generality that $T=1/K$ for $K\in\N$, and hence that $\Omega=1/LT=K/L$ is also
rational.  Indeed, if $T=p/q$ and if for some $L\in\N$, $g=\sum c_n\delta_{nT}$ ($c_n$ with period $L$) identifies $OPW^2(S)$ then letting
$T'=1/q$, $L'=pL$, and $d_n=c_{n/p}$ if $p$ divides $n$ and zero otherwise, then $d_n$ has period $L'$ and $g=\sum_n d_n\delta_{nT'}$.
Note that the set of discontinuities of the function $\CHI_{S+(kT,\ell\Omega)}$, $(k,\ell)\in\Z^2$ in the rectangle $R=[0,T]\times[0,\Omega]$
must occur on line segments of slope $\frac{\sqrt{2}}{2}$ or $\frac{\sqrt{2}}{2} + \frac{1}{4}$ passing through $R$ (that is, intersecting two
edges of $R$).  In order that $\sum \CHI_{S+(kT,\ell\Omega)}$ be constant and hence continuous on $R$, each such segment must coincide with at
least one $(kT,\ell\Omega)$-shift of a different such segment.  In particular, the segment of slope $\frac{\sqrt{2}}{2} + \frac{1}{4}$ containing $(0,0)$
and intersecting one side of $R$ must be met by some $(kT,\ell\Omega)$-shift of the segment joining $(2,\sqrt{2})$ and $(4,2\sqrt{2}+1/2)$, which
implies that this segment must contain a point of the form $(kT,\ell\Omega)$.  However, a simple calculation shows that since $T$ and $\Omega$ are
rational, this is impossible.

Now assume that $T$, and hence also $\Omega=1/(LT)$ is irrational. In this case, discontinuities of $\CHI_{S+(kT,\ell\Omega)}$,
in the rectangle $R$ must lie on lines passing through $R$ with slopes as above, or on a pair of line segments of those slopes terminating at
their intersection point in the interior of $R$ (see Figure~\ref{fig:blowup}(b)).  There are at least one and at most three shifts with
discontinuities of the latter type.  To see this, note that since $T$ is irrational, neither $(2,\sqrt{2})$ nor $(2,\sqrt{2}+1/2)$ lies on a
vertical grid line of the form $t=mT$ and that since $\Omega$ is irrational at least one of these points does not lie on a horizontal
grid line of the form $\nu=n\Omega$.  Similarly, $(4,2\sqrt{2}+1/2)$ cannot lie on a vertical grid line but may lie on a horizontal grid line.
Then by considering the cases in which exactly $1$, $2$, or $3$ of these points do not lie on a horizontal grid line, it is clear that in order
for all discontinuities to be resolved, $(4,2\sqrt{2}+1/2)$ must differ from either $(2,\sqrt{2})$ or $(2,\sqrt{2}+1/2)$ by some $(kT,\ell\Omega)$,
which is impossible since $T$ is irrational.
\end{IEEEproof}

\subsection{Proof of Theorem~\ref{thm:sufficientsamplingrate}.}

\begin{IEEEproof}
Let $L\ge N^2$ be prime, let $\Omega = 1/(TL)$ and let $R_{q,m}$, $(q,m)\in\Gamma$ be the rectangles in the
$(T,N)$-rectification of $S$.  Then letting
$$R'_{q',m'} = [0,T]{\times}[0,\Omega]+(q'T,m'\Omega),$$
$(q',m')\in\Z^2$, each rectangle $R_{q,m}$ is covered by a collection
of rectangles $R'_{q',m'}$ satisfying
$$\sum_{\set{(q',m')\colon R_{q',m'}\cap R_{q,m}\ne\emptyset}}|R_{q',m'}| \le |R_{q,m}| + \frac{2}{L}.$$
Let $\Gamma'$ be those $(q',m')\in\Z^2$ such that $R'_{q',m'}$ has nonempty intersection with $S^\circ$.  Therefore,
\begin{align*}
& \sum_{(q',m')\in\Gamma'}|R'_{q',m'}| \\
& \le  \sum_{(q,m)\in\Gamma}
        \sum_{\set{(q',m')\colon R'_{q',m'}\cap R_{q,m}\ne\emptyset}}|R'_{q',m'}| \\
& \le  \sum_{(q,m)\in\Gamma}  \left(|R_{q,m}| + \frac{2}{L}\right) \\
& \le  \sum_{(q,m)\in\Gamma} |R_{q,m}| + \frac{2N}{L} \\
& \le \frac{\abs{\Gamma}}{N} + \frac{2N}{L} \le \frac{\abs{\Gamma}+2}{N} < |S|(1+\epsilon).
\end{align*}

%

%
%

Consequently, $\abs{\Gamma'}/L\le |S|(1+\epsilon)$, and $S^\circ\subseteq \cup_{(q',m')\in\Gamma'} R'_{q',m'} = R$.
By Theorem~\ref{thm:smallsparksampta13}, we can choose $c\in \C^L$ such that $\norm{c}_0\le\abs{\Gamma'}$,
$Spark(G(c))=\abs{\Gamma'}$ and $c$ is supported on its first $\norm{c}_0$ indices.
Since $S^\circ\subseteq R$, any identifier of $OPW^2(R)$ is also an identifier of $OPW^2(S)$.
Since $R$ consists of only $\abs{\Gamma'}$ rectangles, it follows that vector on the right side of (\ref{eqn:basiclinearsystem})
has at most $\abs{\Gamma'}$ nonzero entries and hence is solvable as long as $Spark(G(c))=\abs{\Gamma'}$.  From this
it follows immediately that $\sum_n c_n\delta_{nT}$ identifies $OPW^2(R)$ and
$$\frac{\norm{c}_0}{L} \le \frac{\abs{\Gamma'}}{L}< |S|(1+\epsilon).$$
\end{IEEEproof}

\subsection{Proof of Theorem~\ref{thm:unknownDomain} and Theorem~\ref{thm:characterizationunknownsupport}}\label{section:proofUnknownDomain}

\begin{IEEEproof} (Theorem~\ref{thm:unknownDomain})
By Theorem~\ref{thm:rectification2}, we can choose $L\in\N$ so that every operator in
${\cal H} (A,B,U,N,\epsilon,1/2)$ has the property that  $supp\,  \eta$ touches at most $L/2$ sets of the form
\begin{align}\label{eqn:sqrtsupport}
 R_{q,m}=[0,1/\sqrt{L}]\times  [0,1/\sqrt{L}] + (q/\sqrt{L},m/\sqrt{L}),
\end{align}
$q,m=-(L-1)/2,-(L-1)/2+1,\ldots, (L-1)/2$.

Now, let $\{S_m:\ m=1,\ldots, \choo{L^2}{L}\}$ be the collection of area 1 sets that are formed by exactly
$L$ subsets of the form $R_{q,m}$ in \eqref{eqn:sqrtsupport}.
Choosing $c\in\C^L$ so that $G(c)$ is full spark, it follows that for each $m$, $OPW(S_m)$ is identifiable with identifier
$\sum_{n\in\Z} c_n\,\delta_{n\sqrt{L}}$ and that constants $C_1, C_2>0$ exist such that
$$C_1\|H\|_{HS}\leq  \|H\sum_{n\in\Z}c_n \delta_{n/\sqrt{L}}\|_{L^2} \leq C_2\|H\|_{HS},$$
for all
$$H\in \bigcup_{m=1,\ldots, \choo{L^2}{L}}OPW^2(S_m).$$
The proof is complete by observing that for $H_1, H_2\in {\cal H} (A,B,U,N,\epsilon,1/2)$ (which is not a linear space), we have $H_1-H_2\in OPW^2(S_m)$ for some $m$, and, hence,
\begin{align*}
C_1\|H_1-H_2\|_{HS} & \leq  \|(H_1 -H_2)\sum_{n\in\Z}c_n \delta_{n/\sqrt{L}}\|_{L^2} \\
                  & \leq C_2\|H_1-H_2\|_{HS},
\end{align*}
$H_1,H_2 \in {\cal H} (A,B,U,N,\epsilon,1/2)$.
Clearly, this leads also to the weaker statement $(H_1-H_2) \sum_n c_n \delta_{n/\sqrt{L}}=0$ implies $H_1=H_2$.
\end{IEEEproof}

\begin{IEEEproof} (Theorem~\ref{thm:characterizationunknownsupport})
The proof of this result follows the proof of Theorem~\ref{thm:characterization}.

\noindent  (i)$\Longrightarrow$(iii) Note first that if $\Delta < 1/2 + 1/(2L)$ then $\Delta L <(L+1)/2$.
Hence if $H_1,\,H_2\in {\mathcal H}_{T,L}(\Delta)$, then with
$\supp\eta_{H_1-H_2}\subseteq \supp\eta_{H_1} \cup \supp\eta_{H_2}=S$,
$$\sum_{k,\ell} \chi_{S+(kT,\ell/(TL))}\leq 2\Delta\,L < L+1$$
and since the left side of the inequality is an integer,
$H_1-H_2\in {\mathcal H}_{T,L}(1)$.  Therefore, \eqref{eqn:characterization1} and \eqref{eqn:characterization2}
hold, and by the same argument as in the proof of Theorem~\ref{thm:characterization} (iii) holds.

\noindent (iii)$\Longrightarrow$(ii)  Obvious.

\noindent (ii)$\Longrightarrow$(i)  Suppose that $\Delta\ge 1/2 + 1/(2L)$.  Then we can find disjoint sets $S_1,\,S_2\subseteq R$ such that
$$ \bigg\|\sum_{k,\ell} \chi_{S_i+(kT,\ell/(TL))}\bigg\|_\infty=\Delta L\ge (L+1)/2$$
This is easily seen by considering the sets
$$\bigcup_{k,\ell}[R_{q,m}+(kTL,\ell/T)]\cap R$$
where for each $0\le q,m<L$,
$$R_{q,m}=([0,T]\times[0,1/TL])+(qT,m/TL).$$
Then $S_1$ and $S_2$ can be formed by choosing two disjoint collections of $\lceil(L+1)/2\rceil$ such sets. Since
$$ \sum_{k,\ell} \chi_{(S_1\cup S_2)+(kT,\ell/(TL))}\ge (L+1)\quad a.e.$$
the same argument as in the proof of Theorem~\ref{thm:characterization} allows us to define distinct operators
$H_1,\,H_2\in {\mathcal H}_{T,L}(\Delta)$
with $\supp\eta_{H_1}\subseteq S_1$ and $\supp\eta_{H_2}\subseteq S_2$ such that $(H_1-H_2)g=0$.
Hence (ii) fails to hold.
\end{IEEEproof}

\section{Conclusion}

This paper contains results relevant to two questions on the identification and recovery of operators with bandlimited symbols
from the response of the operator to a regular weighted delta train.  Such operators model time-variant linear
communication channels.  When the identifier is a weighted delta train, we refer to this the identification as operator sampling
and when the weighting sequence is periodic as regular operator sampling
The procedure is a generalization of classical sampling results for bandlimited functions, and
of the determination of a time-invariant communication channel by measuring its response to
a unit impulse.

We obtain a simple condition on the set $S$ that characterizes when $OPW^2(S)$ can be identified by regular operator sampling.
The condition requires that $S$ be contained in a fundamental domain of a rectangular lattice and that its periodization on a
reciprocal lattice be bounded above by a constant depending on the lattice.  In this case, $|S|\le 1$, and we obtain explicit reconstruction
formulas for the operators in $OPW^2(S)$.
We consider the case in which $S$ is contained in a fundamental domain of a general symplectic lattice and give sufficient
conditions on the lattice under which $OPW^2(S)$ can be identified by regular operator sampling and obtain explicit
reconstruction formulas in this case as well.  We provide an example of a
set $S$ for which $OPW^2(S)$ can be identified by operator sampling but not by regular operator sampling.

For these results it is
required that the support set be known.  We also obtain a result showing that, under mild geometric conditions,
recovery is possible when the support set is unknown but has area smaller than $1/2$ and we characterize all support sets
for which identification is possible via regular operator sampling when the support set has area $\le 1/2$.
This characterization allows us to define a large class of operators for which identification is possible when the spreading
support is small.  This class includes the class similarly characterized in \cite{BH11, BH13}.
It is shown in \cite{BH11, BH13} that this class can be identified without knowledge of the spreading support
for areas less than one.  Following the ideas given in \cite{BH11, BH13}, we define a larger class of operators with area less than one that can
be similarly identified without knowing the spreading support.

Finally, we give a necessary condition on the rate of sampling, that is, the average number of deltas in the identifying weighted
delta train per unit time, required to identify an operator with bandlimited symbol.  The necessary rate depends on the bandwidth
of the spreading support.  We give a sufficient condition on the sampling rate in terms of the area of the spreading region.
As a consequence of this result, it is observed that if the area of the spreading support is small, then any operator in the
class of operators having that spreading support can be identified by only a portion of its response to an appropriate identifier.
The fraction of the response sufficient for identification is asymptotically proportional to the area of the spreading support.

\def\cprime{$'$} \def\cprime{$'$}

\begin{IEEEbiographynophoto}{G\"otz~E.~Pfander}
received his Ph.D.\ degree in mathematics from the University of
Maryland in 1999. Since 2002 he has been Associate Professor of
Mathematics at Jacobs University Bremen, Germany.
\end{IEEEbiographynophoto}

\begin{IEEEbiographynophoto}{David~F.~Walnut}
received his Ph.D.\ degree in mathematics from the University of
Maryland, USA in 1989. He is Professor of Mathematics at George
Mason University, Virginia, USA.
\end{IEEEbiographynophoto}

\end{document}